\documentclass[aps,a4paper, preprint, superscriptaddress,preprintnumbers,floatfix,nofootinbib,amsmath,amssymb,table]{revtex4-1}
\usepackage{hyperref}
\usepackage{cleveref}
\usepackage{subfig}

\usepackage{amstext,amssymb}
\usepackage[section]{placeins}
\usepackage{placeins}
\usepackage{amsmath}
\usepackage{graphicx}
\usepackage{xspace}
\usepackage{color}
\usepackage{float}
\usepackage{comment}
\usepackage{amstext,amssymb}
\usepackage{amsmath}
\usepackage[section]{placeins}
\usepackage{placeins}
\usepackage{xspace}
\usepackage{units}
\usepackage{array}
\usepackage{amsmath,bm}
\usepackage{amssymb}
\usepackage{gensymb}
\usepackage{soul}
\usepackage{textcomp}
\usepackage{slashed} 
\usepackage{multirow}
\usepackage{hyperref}
\usepackage{mathrsfs}
\usepackage{enumitem}
\usepackage{graphicx}
\usepackage{comment}
\usepackage{epstopdf}
\usepackage{float}
\usepackage{units}
\usepackage{feynmp}
\usepackage{slashed}
\usepackage{amsmath,bm}
\usepackage[table]{xcolor}
\usepackage[normalem]{ulem}
\definecolor{antiquewhite}{rgb}{0.98, 0.92, 0.84}
\definecolor{aqua}{rgb}{0.0, 1.0, 1.0}
\definecolor{amethyst}{rgb}{0.6, 0.4, 0.8}
\definecolor{applegreen}{rgb}{0.55, 0.71, 0.0}
\definecolor{byzantine}{rgb}{0.74, 0.2, 0.64}
\definecolor{cadetgrey}{rgb}{0.57, 0.64, 0.69}
\definecolor{candypink}{rgb}{0.89, 0.44, 0.48}
\usepackage{lipsum}
\usepackage{cals}
\usepackage{slashed} 
\newcommand{\be}{\begin{equation}}
\newcommand{\ee}{\end{equation}}
\newcommand{\bea}{\begin{eqnarray}}
\newcommand{\eea}{\end{eqnarray}}
\newcommand{\nn}{\nonumber}

\def\s1{\hat s}

\makeatletter

    \def\CT@@do@color{%
      \global\let\CT@do@color\relax
            \@tempdima\wd\z@
            \advance\@tempdima\@tempdimb
            \advance\@tempdima\@tempdimc
    \advance\@tempdimb\tabcolsep
    \advance\@tempdimc\tabcolsep
    \advance\@tempdima2\tabcolsep
            \kern-\@tempdimb
            \leaders\vrule
                    \hskip\@tempdima\@plus  1fill
            \kern-\@tempdimc
            \hskip-\wd\z@ \@plus -1fill }
    \makeatother

\usepackage{slashed}
\definecolor{ashgrey}{rgb}{0.7, 0.75, 0.71}
\definecolor{aureolin}{rgb}{0.99, 0.93, 0.0}
\definecolor{babypink}{rgb}{0.96, 0.76, 0.76}
\definecolor{buff}{rgb}{0.94, 0.86, 0.51}
\definecolor{chamoisee}{rgb}{0.63, 0.47, 0.35}
\definecolor{chartreuse(web)}{rgb}{0.5, 1.0, 0.0}
\definecolor{citrine}{rgb}{0.89, 0.82, 0.04}
\definecolor{emerald}{rgb}{0.31, 0.78, 0.47}
\definecolor{fawn}{rgb}{0.9, 0.67, 0.44}
\definecolor{fulvous}{rgb}{0.86, 0.52, 0.0}

\newcommand{\nua}[1]{\ensuremath{\rlap{\kern-2.5pt\ensuremath{\overset{\scriptscriptstyle(-)}{\phantom{\nu}}}}{\ensuremath{{\nu}_{#1}}}}\xspace}

\begin{document}
\title{Probing  neutrino mass ordering with supernova neutrinos at NO$\nu$A including the effect of sterile neutrinos
}
\author{Papia Panda}
\email{ppapia93@gmail.com}
\affiliation{School of Physics,  University of Hyderabad, Hyderabad - 500046,  India}
\author{Rukmani Mohanta}
\email{rmsp@uohyd.ac.in}
\affiliation{School of Physics,  University of Hyderabad, Hyderabad - 500046,  India}

\begin{abstract}
In this work, we explore the possibility of probing the mass ordering sensitivity as a function of supernova distance in the context of the ongoing neutrino experiment NO$\nu$A. We provide a detailed study of the active-active and active-sterile mixing frameworks, illustrating how supernova neutrinos can be used to realize the existence of sterile neutrinos.  
Interestingly, we infer that observation of the NC channel alone can differentiate between the presence and absence of sterile neutrinos. 
Our results indicate that the primary channel of NO$\nu$A can distinguish normal mass ordering from inverted mass ordering at  $5 \sigma$ confidence level for a supernova explosion occurring at a distance of 5 kpc. Additionally, we examine the impact of systematic uncertainties on mass ordering sensitivity, showing that higher levels of systematics lead to a reduction in sensitivity. Similarly, the inclusion of energy smearing  significantly diminishes ordering sensitivity.

\end{abstract}
\maketitle

\section{Introduction}
The core of a massive star with a mass greater than $8 M_{\odot}$, where $M_{\odot}$ is the mass of the sun, collapses with a tremendous amount of energy and light at the end of its life, producing a ``core-collapse supernova". Approximately $99\%$ of this energy is carried away by neutrinos of various types, and 
their weakly interacting nature provides valuable insights into the supernova explosion mechanism.

The phenomenon of neutrino oscillation confirms that neutrinos possess non-zero masses. The flavor ($\nu_e$, $\nu_{\mu}$, $\nu_{\tau}$) and mass  eigenstates ($\nu_1$, $\nu_2$, $\nu_3$) of neutrinos are related by the unitary PMNS matrix, comprising of three mixing angles and one CP violating phase. Consequently, neutrino oscillation depends on six parameters: $\theta_{12}$, $\theta_{23}$, $\theta_{13}$, $\Delta m_{21}^2$, $\Delta m_{31}^2$, and $\delta_{CP}$. Here, $\Delta m_{ij}^2 = m_i^2 - m_j^2$, and $\delta_{CP}$ is the CP-violating phase. These mixing parameters are determined very precisely except the CP phase $\delta_{CP}$,  the sign of $\Delta m_{31}^2$,  and the octant of atmospheric mixing angle $\theta_{23}$. A positive $\Delta m_{31}^2$ indicates normal mass ordering (NO), while a negative sign implies inverted ordering (IO).
In addition to long-baseline and reactor neutrino experiments, supernova neutrinos also present a highly effective option for addressing the mass ordering problem. This study aims to investigate the neutrino mass ordering problem using supernova neutrinos.

Some short-baseline neutrino experiments \cite{MiniBooNE:2018esg,MiniBooNE:2020pnu, LSND:2001aii,IceCube:2016rnb} hint towards the existence of additional neutrino flavors, referred to as sterile neutrinos. However, experimental confirmation is not yet established. Sterile neutrinos, unlike active neutrinos, do not directly interact with Standard Model (SM) particles. Interactions among active neutrinos are described as the active-active framework, whereas interactions involving sterile neutrinos form the active-sterile framework. Information on the production, propagation, and detection of sterile neutrinos is limited, and one promising source for studying them is core-collapse supernovae. Several studies \cite{Saez:2021fpa,Esmaili:2014gya,Boyarsky:2009ix,Tamborra:2011is,Wu:2013gxa,Collin:2016aqd,Yudin:2016zqp,Franarin:2017jnd,Qian:2018ngd,Saez:2018zsb,Tang:2020pkp,Saez:2021rxq,Fushimi:2021kce,Fetter:2002xx,Tamborra:2012iz} have explored sterile neutrinos in supernovae, with some considering their presence at the core and others at the oscillation level. In this study, we assume that sterile neutrinos are produced during neutrino oscillations occurring in the region between the core and the surface of the supernova.

NO$\nu$A (NuMI Off-axis $\nu_e$ Appearance) is a currently running long-baseline  experiment which has the potential to detect supernova neutrinos during its operational period. Several studies \cite{NOvA:2020dll,Vasel:2021ymu,NOvA:2021zhv} have simulated supernova neutrino events at NO$\nu$A. The motivation for this work stems from the possibility of a supernova explosion occurring within the next few years. In such a scenario, the currently running long-baseline neutrino experiments, such as NO$\nu$A or T2K, could extract valuable information from supernova neutrinos. While this study focuses on the NO$\nu$A detector, a similar investigation could also be conducted using the T2K experiment. There are couple of works \cite{Panda:2023rxa,Gaba:2024asp} on the mass ordering with supernova neutrinos in future neutrino experiments like DUNE, T2HK, T2HKK, and JUNO. These upcoming  experiments with their larger detector volumes and higher statistics, are expected to provide deeper insights. However, they are   likely to be in operation within a decade or more. 
Here, we present an analysis of the prospects of observing supernova neutrinos at NO$\nu$A in both the active-active and active-sterile frameworks, considering a single additional sterile neutrino.

The structure of the paper is outlined as follows. Sec. \ref{theory} provides a brief overview of the theory of supernova neutrino oscillations in both active-active and active-sterile frameworks. Sec. \ref{experi} details the experimental setup of the NO$\nu$A detector and the simulation methodology employed in this study. In Sec. \ref{main-channel}, the primary detection channels for supernova neutrinos are discussed. Sec. \ref{event-rate-section} presents the event rate calculations for various channels under both the frameworks. The key findings of our study are summarized in Sec. \ref{results}, which is divided into three subsections: mass ordering sensitivity analysis, the impact of systematic uncertainties, and the influence of energy smearing on mass ordering sensitivity. Finally, Sec. \ref{concluding remarks} concludes with our observations and remarks. The event rates for both active-active and active-sterile scenarios are provided in Appendix \ref{app-a}.

\section{Theoretical formalism}
\label{theory}

During the initial phase of a supernova burst, electron-type neutrinos ($\nu_e$) dominate, as they are primarily produced through electron capture on protons and nuclei when the neutrinosphere is affected by the shock wave \cite{Mirizzi:2015eza}. These electron neutrinos interact strongly with matter, resulting in a lower average energy for $\nu_e$ compared to other neutrino types ($\nu_{\mu}, \nu_{\tau}$).  Similarly, $\bar{\nu}_e$ neutrinos, which interact with matter via charged current interactions, also exhibit relatively low average energy, although it remains higher than that of $\nu_e$. 
On the other hand, non-electron type neutrinos ($\nu_{\mu}, \nu_{\tau}$), commonly referred to as $\nu_x$, are unable to interact via charged current at the energies of supernova neutrinos (a few MeV).
These neutrinos interact only through neutral currents, resulting in the highest average energy for $\nu_x$.  In the primary neutrino spectra, the average energy hierarchy is expected to follow the relation \cite{Dighe:1999bi}:  
\begin{equation}
    \langle E^0_{\nu_e} \rangle
    < \langle E^0_{\bar{\nu}_e} \rangle < \langle E^0_{\nu_x} \rangle.
\end{equation}
The flavor dependent primary neutrino spectra of the supernova neutrinos at the core can be expressed by power law distribution,
\begin{equation}
    \Phi_{\nu} (E_{\nu}) = \mathcal{N} \left( \frac{E_{\nu}}{\langle E_{\nu} \rangle} \right)^{\alpha} e^{-(\alpha+1) \frac{E}{\langle E  _{\nu} \rangle}}\;,
\end{equation}
where $E_{\nu}$ and $\alpha$ are the neutrino energy and pinching parameter respectively. $\mathcal{N}$ is the normalization constant with the expression,
\begin{equation}
    \mathcal{N}= \frac{(\alpha+1)^{\alpha+1}}{\langle E_{\nu} \rangle \Gamma(\alpha+1)}\;.
    \label{nor}
\end{equation}
The flux at the core of the supernova can be written in terms of flavor dependent primary neutrino spectra $\Phi_{\nu}(E_{\nu})$, average neutrino energy $\langle E_{\nu} \rangle$ and luminosity $L_{\nu}$ by the relation,
\begin{equation}
    F_{\nu}^0 = \frac{L_{\nu}}{\langle E_{\nu} \rangle} \Phi_{\nu} (E_{\nu})\;.
    \label{flux}
\end{equation}
In presence of Mikheyev-Smirnov-Wolfenstein (MSW) \cite{Wolfenstein:1977ue,Mikheev:1987qk} effect, the oscillated flux has  different expression for active-active and active-sterile frameworks. In our simulation, we took all three supernova neutrino development stages and
integrated them over time to obtain fluence ($F_{\nu}^0$).

In the following subsections, we discuss the oscillated neutrino flux for these two neutrino frameworks.

\subsection{Active-active neutrino framework}
The fluence at the detector on Earth in the active-active framework is expressed as:
\begin{eqnarray}
    &&F_{\nu_e} = p F_{\nu_e}^0 + (1-p) F_{\nu_x}^0 \;,\nn\\
   && F_{\bar{\nu}_e} = \bar{p} F_{\bar{\nu}_e}^0 + (1- \bar{p} ) F_{\nu_x}^0 \;,\nn\\
&&   2F_{\nu_x} = (1-p)F_{\nu_e}^0 + (1+p) F_{\nu_x}^0\;, \nn \\
&& 2F_{\bar{\nu}_x} = (1-\bar{p})F_{\bar{\nu}_e}^0 + (1+\bar{p}) F_{\bar{\nu}_x}^0\;,
\label{3-nu-exp-exp}
\end{eqnarray}
where $p$ and $\bar{p}$
  represent the survival probabilities for electron neutrinos and electron antineutrinos, respectively and the corresponding expressions are provided in Table \ref{3-nu-exp}.
  From the Table, we observe that
$p$ and $\bar{p}$
  have different values for normal and inverted mass orderings, which can thus yield a non-zero sensitivity to the mass ordering in this framework. One can further  notice that the probabilities depend on the solar and reactor mixing angles, with no dependence on the atmospheric mixing angle $\theta_{23}$. As a result, the fluence at the detector varies according to these two angles.
Following Ref.~\cite{Dighe:1999bi}, we have assumed the adiabatic condition for all our calculations, meaning that both the high-density and low-density flip probabilities, 
$P_H$
  and $P_L$, are zero inside the supernova.

\begin{table}[htbp]
    \centering
    \begin{tabular}{|c|c|c|}
    \hline
    \rowcolor{citrine!10}
        Ordering & $p$ & $\bar{p}$  \\
        \hline
        \rowcolor{babypink!50}
        Normal & $\sin^2 \theta_{13}$ & $\cos^2 \theta_{12} \cos^2 \theta_{13}$  \\
        \hline
        \rowcolor{aureolin!30}
         Inverted & $\sin^2 \theta_{12} \cos^2 \theta_{13}$  &  $\sin^2 \theta_{13}$\\
        \hline
    \end{tabular}
    \caption{In active-active neutrino framework, survival probability expressions of neutrino ($p$) and antineutrino ($\bar{p}$) fluxes for two cases: normal ordering and inverted ordering.}
    \label{3-nu-exp}
\end{table}

While using the neutronization burst could enhance mass ordering sensitivity, our analysis relies on the total time-integrated fluence due to NO$\nu$A’s detector constraints. Since NO$\nu$A lacks the millisecond-scale timing resolution and low-energy triggering efficiency required to resolve the neutronization burst separately, we adopt a fluence-based approach to maximize detection statistics over the entire $\sim$10 s burst. Future studies could explore the feasibility of a dedicated neutronization burst analysis, possibly with real-time triggering improvements in NO$\nu$A or next-generation detectors.
The fluence for different flavor neutrinos from Eqn.(\ref{3-nu-exp-exp}) are depicted graphically in Figure \ref{flux}. To generate the panels of Figure \ref{flux}, we use the values of the oscillation parameters as  listed in Table \ref{osc-params} and  have considered Garching electron capture supernova model \cite{Hudepohl:2009tyy} of our simulation. We checked that the conclusion of Figure \ref{flux} remains valid for any choice of oscillation parameters within the $3\sigma$ allowed region. In this figure, the left (right) panel of upper row shows the time-integrated flux, or fluence, as a function of neutrino energy for $ \nu_e  ( \bar{\nu}_e)$ across five different cases. Similarly, the left (right) panel of lower row illustrates the fluence of $\nu_x (\bar{\nu}_x)$ under varying conditions. In each panel, the red dashed-dotted curve represents the unoscillated data, while the magenta (cyan) curve shows the fluence for normal (inverted) ordering in the active-active neutrino framework.

\begin{table}[]
    \centering
    \begin{tabular}{||c||c||}
    \hline
    \hline
       Oscillation parameters  & Values \\
       \hline
       \hline
       $\theta_{12}$  & $33.41^{\circ}$\\
       \hline
       \hline
       $\theta_{13}$  &  $8.58^{\circ}$  \\
       \hline
       \hline
       $\theta_{23}$  &  $42.20^{\circ}$  \\
       \hline
       \hline
       $\theta_{14}$ &  $5^{\circ}$  \\
       \hline
       \hline
       $\Delta m_{21}^2$  &  $7.410 \times 10^{-5}$ ${\rm eV}^2$  \\
       \hline
       \hline
       $\Delta m_{31}^2$  &  $ \pm 2.507 \times 10^{-3}$ ${\rm eV}^2$  \\
       \hline
       \hline
       $\Delta m_{41}^2$   &  $1$ ${\rm eV}^2$  \\
       \hline
       \hline
    \end{tabular}
    \caption{Neutrino oscillation parameter values \cite{Esteban:2020cvm} used in the study.}
    \label{osc-params}
\end{table}

From the left panel of the upper row, we observe that the red dashed-dotted curve has a significantly higher fluence than the cyan and magenta solid curves. Between the cyan and magenta curves, the cyan curve is larger, indicating that for $\nu_e$ fluence, the inverted ordering (IO) results in higher fluence compared to the normal ordering (NO). 
Conversely, for the $\bar{\nu}_e$ case (right panel of the upper row), all cases appear to closely overlap, making it challenging to distinguish between them.
In the left panel of the bottom row, unlike the $\nu_e$ distribution, the red dash-dotted curve has a lower fluence compared to the cyan and magenta solid curves. Compared to the cyan curve, the magenta one is higher, showing that for $\nu_x$ fluence, the NO results in higher fluence than the IO.
For the fluence of $\bar{\nu}_x$, similar to the that of $\bar{\nu}_e$, all the curves overlap, making it difficult to distinguish between them. 
The behavior observed in Figure \ref{flux} results from the interplay of two mixing angles, $\theta_{12}$ and $\theta_{13}$, as well as the initial fluences of the different neutrino flavors.

\begin{figure}
\includegraphics[scale=0.5]{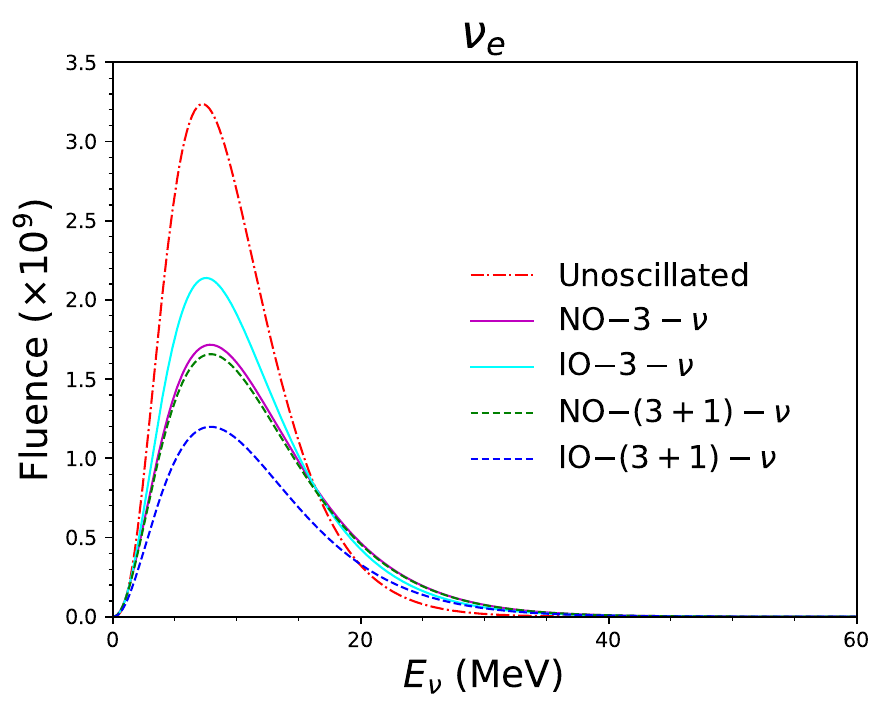}
\includegraphics[scale=0.5]{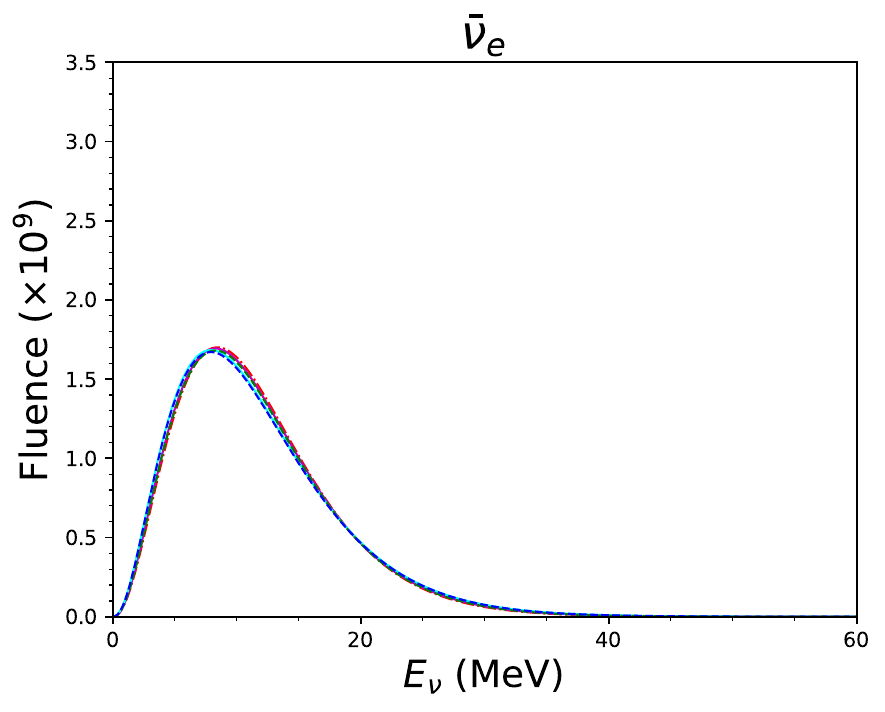}
    \\
\includegraphics[scale=0.5]{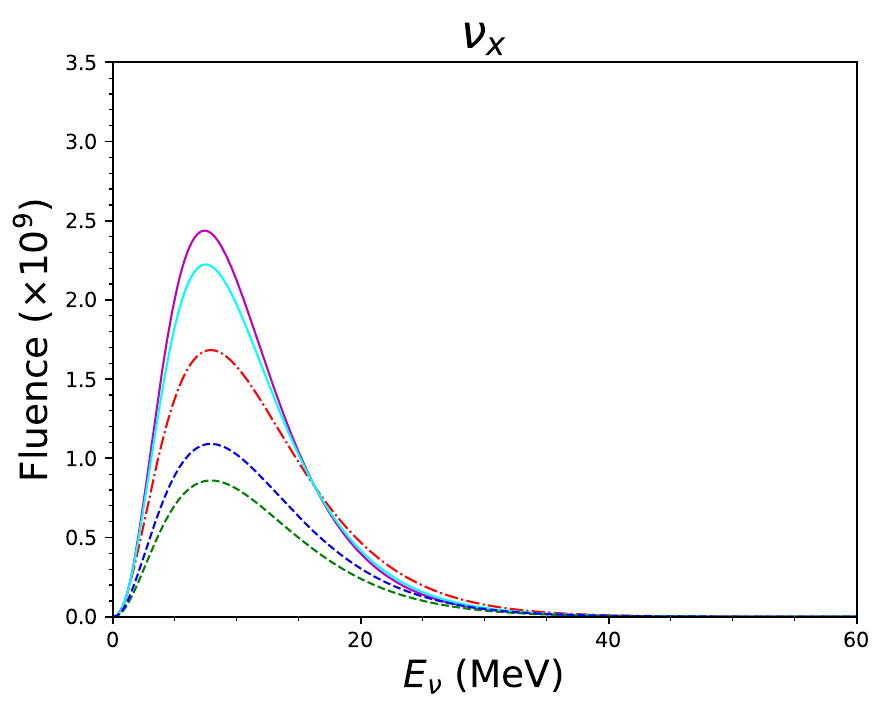}
\includegraphics[scale=0.5]{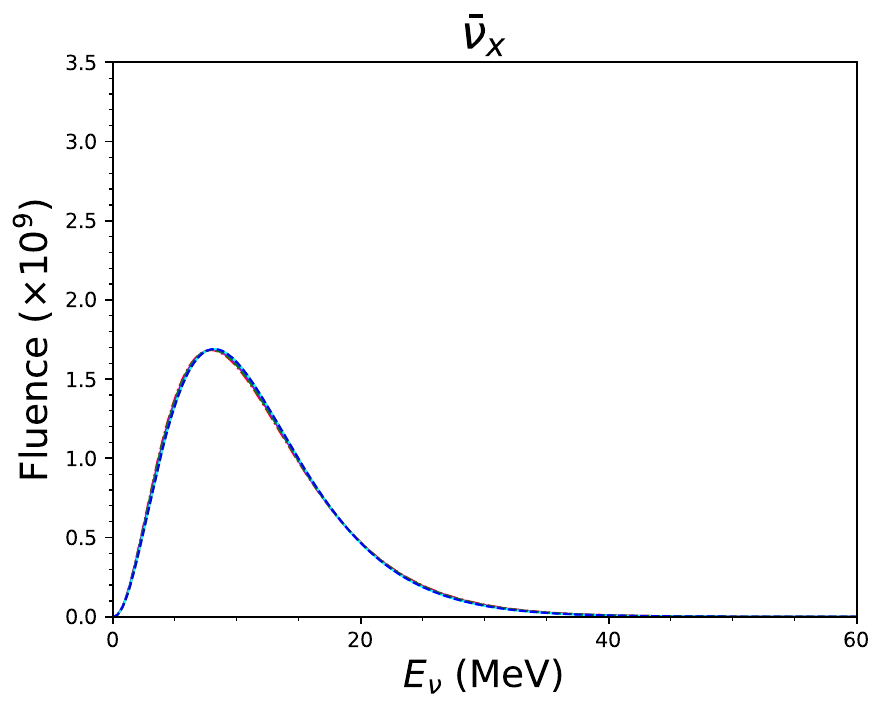}
    \caption{Fluence (integrated flux over time) as a function of neutrino energy ($E_{\nu}$) in MeV.  Left (right) of upper row is for $\nu_e (\bar{\nu}_e)$ while left (right) of lower panel is for $\nu_x (\bar{\nu}_x)$ flavor. In each panel, color codes are given in the legend. }
    \label{flux}
\end{figure}
\subsection{Active-sterile neutrino framework}
For active-sterile scenario, the expression for active and sterile neutrino fluxes are as follows \cite{Saez:2021fpa}
\begin{eqnarray}
    F_{\nu_e} &=& a_{ee} F_{\nu_e}^0 + a_{e x} F_{\nu_x}^0 + a_{e s} F_{\nu_s}^0 ,\nonumber \\
    F_{\bar{\nu}_e}&=& b_{ee} F_{\bar{\nu}_e^0} + b_{ex} F_{\bar{\nu}_x^0} + b_{e s} F_{\bar{\nu}_s}^0, \nonumber \\
    2 F_{\nu_x} &=& (a_{\mu e } + a_{\tau e } ) F_{\nu_e}^0 + (a_{\mu x} + a_{\tau x}) F_{\nu_x}^0  + 
 (a_{\mu s} + a_{\tau s}) F_{\nu_s}^0,
 \nonumber \\
    2 F_{\bar{\nu}_x} &=& (b_{\mu e} + b_{\tau e} ) F_{\bar{\nu}_e}^0 + (b_{\mu x} + b_{\tau x}) F_{\bar{\nu}_x}^0  + (b_{\mu s} + b_{\tau s}) F_{\bar{\nu}_s}^0 ,\nonumber \\
    F_{\nu_s} &=& a_{s e} F_{\nu_e}^0 + a_{s x} F_{\nu_x}^0  + a_{s s} F_{\nu_s}^0,  \nonumber \\
    F_{\bar{\nu}_s} &=& b_{s e } F_{\bar{\nu}_e}^0 + b_{s x} F_{\bar{\nu}_x}^0  + b_{s s} F_{\bar{\nu}_s}^0 ,
    \label{3+1-neu-exp}
 \end{eqnarray}
where the expressions of $a_{\alpha e}$, $a_{\alpha x}$, $b_{\alpha e}$ and $b_{\alpha x}$ for normal order are given as
 \begin{eqnarray}
 a_{\alpha e} &=& |U_{\alpha 1}|^2 P_H P_L (1-P_S) + |U_{\alpha 3}|^2 P_S + |U_{\alpha 2}|^2 P_H (1-P_L) (1-P_S), \nonumber \\
 &+& |U_{\alpha 4}|^2 (1-P_S) (1-P_H) , \nonumber \\
 a_{\alpha x}& = &|U_{\alpha 1}|^2 (1-P_H P_L) + |U_{\alpha 2}|^2 (1-P_H + P_H P_L) + |U_{\alpha 4}|^2 P_H, \nonumber \\
b_{\alpha e} &=& |U_{\alpha 1}|^2 \nonumber \\
b_{\alpha x} &=& |U_{\alpha 2}|^2 + |U_{\alpha 3}|^2.
\label{ae-nh}
 \end{eqnarray}
For the inverted ordering case, the expressions are,
 \begin{eqnarray}
     a_{\alpha e} &=& |U_{\alpha 1}|^2 P_L (1-P_S) + |U_{\alpha 2}|^2 P_S + |U_{\alpha 4}|^2 (1-P_S) (1-P_L), \nonumber \\
     a_{\alpha x} &=& |U_{\alpha 1}|^2 (1-P_L) + |U_{\alpha 3 }|^2 + |U_{\alpha 4}|^2 P_L, \nonumber \\
     b_{\alpha e} &=&|U_{\alpha 2}|^2 \bar{P}_H + |U_{\alpha 3}|^2 (1-\bar{P}_H), \nonumber \\
     b_{\alpha x} &=& |U_{\alpha 1}|^2 + |U_{\alpha 2}|^2 (1-\bar{P}_H) + |U_{\alpha 3}|^2 \bar{P}_H.
  \end{eqnarray} 
We further assume there are no initial sterile neutrinos at the core of the supernova, so $F_{\nu_s}^0$ and $F_{\bar{\nu}_s}^0$ are zero \footnote{This assumption is very much relevant to our study, as we consider the sterile neutrino mass $m_s \sim {\cal O}(1)$ eV. The production of such a small mass range within the supernova core is suppressed due to strong matter potentials, non-thermal production mechanisms, decoherence effects, and its connections to cosmology and early-universe production.}. 

Similar to active-active case, here also, in adiabatic approximation, survival probabilities of electron (anti)neutrino in high and low densities are set to be zero, i.e., 
  $P_H = \bar{P}_H = P_L = \bar{P}_L = P_S = \bar{P}_S = 0 $. The $4 \times 4$ mixing matrix can be parametrized as $U_{\alpha i} = R_{14}\cdot R_{23} \cdot R_{13}\cdot R_{12}$ ordering, where $R_{i j}$ are the rotation matrices in $i-j$ plane.  To simplify our analysis, we assume the CP violating phases to be zero.
Under the adiabatic condition, the expressions of $a_{\alpha e (x)} $ and $b_{\alpha e (x)}$ are given in Table \ref{3+1 nu exp}. Depending upon the values of the mixing angles and the initial fluence of different flavors,  Eq. \ref{3+1-neu-exp} has different values in normal and inverted scenarios.
\begin{table}[htbp]
    \centering
    \begin{tabular}{|c|c|c|c|c|}
    \hline
    \rowcolor{green!50!yellow!20}
        Ordering & $a_{\alpha e }$ & $a_{\alpha x}$  & $b_{\alpha e}$  & $b_{\alpha x}$ \\
        \hline
        \rowcolor{green!20!yellow!40}
        Normal & $|U_{\alpha 4}|^2$ & $|U_{\alpha 1}|^2 + |U_{\alpha 2}|^2$  & $|U_{\alpha 1}|^2$  &  $|U_{\alpha 2 }|^2 + |U_{\alpha 3}|^2$ \\
        \hline
        \rowcolor{green!10!red!20}
         Inverted & $|U_{\alpha 4}|^2$ & $|U_{\alpha 1 }|^2 + |U_{\alpha 3}|^2$ & $|U_{\alpha 3}|^2$  &  $|U_{\alpha 1 }|^2 + |U_{\alpha 2}|^2$\\
        \hline
    \end{tabular}
    \caption{In the active-sterile neutrino framework, the expressions for the couplings of neutrinos ($a_{\alpha e}, a_{\alpha x}$) and anti-neutrino ($b_{\alpha e}, b_{\alpha x}$) are provided for two scenarios: normal ordering and inverted ordering.}
    \label{3+1 nu exp}
\end{table}
It is noteworthy that,  the coupling of (anti)neutrinos depends on all the mixing angles: $\theta_{12}, \theta_{13}, \theta_{23},$ and $\theta_{14}$, as seen from Table \ref{3+1 nu exp}. Thus, the presence of a sterile neutrino offers the possibility of gaining insight into the octant sensitivity of $\theta_{23}$ through supernova neutrinos. This is in contrast to the active-active framework, where $\theta_{23}$ does not appear in the expressions of $ p$ and $\bar{p}$.

The expressions in eq. \ref{3+1-neu-exp} are graphically represented in Fig.~\ref{flux}. To generate the panels in Fig. \ref{flux}, we use the sterile neutrino oscillation parameters listed in Table \ref{osc-params}. In this figure, the green and blue dashed curves represent the fluence for active-sterile neutrino scenario. The green dashed curve corresponds to the NO, while the blue dashed curve represents the IO.

In the left panel of the upper row, we observe that the red dash-dotted curve is significantly higher than the green and blue dashed curves. Comparing the green and blue curves, we find that, unlike the active-active framework, for active-sterile scenario, the NO spectrum is higher than the IO spectrum. For the $\bar{\nu}_e$ case, all the expressions appear to overlap closely, making them indistinguishable.
In the left panel of the bottom row, similar to the $\nu_e$ distribution, the red dash-dotted curve shows a higher fluence compared to the green and blue dashed curves. When comparing the green and blue curves, the blue curve is higher than the green. For $\bar{\nu}_x$, the blue, green dashed, and red dash-dotted curves overlap, making it difficult to distinguish between the NO and IO results.

It is worth comparing the active-active scenario with the active-sterile framework. The following conclusions can be drawn from Fig. \ref{flux},
\begin{itemize}
    \item For the $\nu_e$ fluence, the active-active scenario is always greater than the active-sterile framework across the entire energy range, regardless of whether the ordering is normal or inverted.
    \item The behavior for the $\bar{\nu}_e$ fluence is exactly the same as that for the $\nu_e$ fluence.
    \item Similarly, the $\nu_x$ fluence exhibits the same behavior as the $\nu_e$ fluence.
    \item For the $\bar{\nu}_x$ case, the active-active  and  active-sterile  scenarios, the fluences completely overlap across the entire energy range.
\end{itemize}

In our calculations, we have considered only the Mikheyev-Smirnov-Wolfenstein (MSW) effect \cite{Dighe:1999bi}. However, within the supernova, numerous flavor-changing processes occur. One significant effect following the MSW effect is the ``collective effect". Extensive studies \cite{Mirizzi:2015eza, Chakraborty:2016yeg, Horiuchi:2018ofe, Tamborra:2020cul} have explored the collective effect in supernova neutrinos, but its outcomes remain largely uncertain. We currently have limited knowledge about the collective effect, including the type of its nature, mechanisms, and behavior. Due to this lack of a clear understanding, we have not included the collective effect in our calculations of fluence for different neutrino flavors.

\section{Experimental setup and simulation details}
\label{experi}

In this section, we discuss the key features of the NO\(\nu\)A detector and the simulation tools utilized in our study. The NuMI Off-axis \(\nu_e\) Appearance (NO\(\nu\)A) experiment \cite{NOvA:2007rmc,NOvA:2016vij,NOvA:2017ohq,NOvA:2017abs,NOvA:2018gge,NOvA:2019cyt} is a long-baseline accelerator neutrino experiment currently in operation. It employs two functionally identical detectors: one near detector and one far detector. Both detectors are constructed using planes of extruded polyvinyl chloride (PVC) with a custom formulation that includes titanium dioxide, with $^{12}\text{C}$ as the primary component \cite{Mufson:2015kga}. 
The fiducial volumes of the near and far detectors are 300 tons and 14 kilotons, respectively. The far detector, being nearly 50 times larger than the near detector, is significantly more sensitive to signals from supernova neutrinos. As a result, we focus on the far detector for the remainder of our simulations.

One of the primary challenges in detecting supernova neutrinos at NO$\nu$A arises from the high rate of cosmic-ray-induced background events.
Unlike underground detectors such as Super-Kamiokande and DUNE, NO$\nu$A's far detector (FD) is located at the surface with a minimal overburden, leading to a background rate of approximately 150 kHz. These backgrounds mainly consist of Cosmic-ray muons producing secondary particles via spallation, Michel electrons from muon decays mimicking low-energy neutrino interactions and  high-energy showers that could obscure supernova signals. 

To mitigate these backgrounds, Ref. \cite{Vasel:2021ymu} suggests several rejection techniques such as hit clustering, near detector (ND) comparisons, and energy thresholding. In our present work, we adopt a conservative approach by explicitly imposing a lower energy threshold of 10 MeV in all our event rate and sensitivity calculations. This threshold follows from the observation in Ref. \cite{Vasel:2021ymu} 
that below 10 MeV, signal discrimination becomes unreliable in the NO$\nu$A far detector.
While this assumption simplifies the analysis, we acknowledge its limitation and emphasize that incorporating realistic background modeling will be an important aspect of future work.

For our simulations, we use the Supernova Neutrino Observatories with GLoBES (SNOwGLoBES) software~\cite{snowglobes}, which is based on the GLoBES framework~\cite{Huber:2004ka,Huber:2007ji}. This tool is specifically designed to study supernova neutrinos. SNOwGLoBES calculates event rates by utilizing input parameters such as neutrino fluxes, cross sections, and detector configurations. For calculating mass ordering sensitivity, we apply the Poisson log-likelihood statistical formula, given by the expression:
\begin{equation}
    \chi^2_{\rm stat} = 2 \sum_{i=1}^n \left[N_i^{\rm test} - N_i^{\rm true} - N_i^{\rm true} \rm{log} \left( \frac{N_i^{ \rm test}}{N_i^{\rm true}} \right) \right],
    \label{chi}
\end{equation}
where, $N_{i}^{\rm true}$ and $N_{i}^{\rm test}$ are the event rates of true and test spectra respectively, with $i$ denoting the energy bin index. In our simulation, the detected energy range spans from 0.5 MeV to 100 MeV. We have divided this range into 200 true energy bins and 200 sampling bins, which corresponds to $n = 200$ bins in eqn \ref{chi}.
To determine $\chi^2$, we use the oscillation parameter values from Table \ref{osc-params} for the true spectrum, while in the test spectrum, we vary $\theta_{23}$ and $\Delta m_{31}^2$ within their $3\sigma$ confidence level ranges, keeping all other parameters fixed.
To evaluate the impact of systematic errors on physics sensitivities, we consider two types of errors: normalization error and energy calibration error. The normalization error affects the overall amplitude of the spectrum while preserving its structure. In contrast, the energy calibration error varies with the energy value of each bin, altering the shape of the event spectrum. When both errors are included, the test event rate is expressed as:
\begin{equation}
    N_i^{\rm test} \rightarrow N_i^{\rm test} [( 1+ a ) + b(E_i^{\prime} - \bar{E^{\prime}})/(E^{\prime}_{\rm max}-E^{\prime}_{\rm min})],
    \label{error}
\end{equation}
Here, $a$ and $b$ are the nuisance parameters corresponding to the normalization and energy calibration errors, respectively. For a $5\%$ systematic error in both types, the nuisance parameters $a$ and $b$ can be expressed in terms of the pull variables $p_1$ and $p_2$ as
\begin{equation}
    a=0.05~ p_1 ,  ~~~~~~ b= 0.05~ p_2.
\end{equation}
Finally, in presence of systematics errors \cite{Gonzalez-Garcia:2004pka}, the final expression of sensitivity is,
\begin{equation}
    \chi^2_{ \rm stat+sys} = \chi^2_{\rm stat} + p_1^2 +p_2^2 \;.
    \label{chi-sys}
\end{equation}
\section{Main Channels}
\label{main-channel}

Since the far detector of NO$\nu$A is a scintillator detector, the primary interaction with supernova neutrinos is through inverse beta decay (IBD). In this process, an electron antineutrino interacts with a proton, resulting in the production of a neutron and a positron, as represented by the following equation,
\begin{eqnarray}
    \bar{\nu}_e + p \rightarrow n + e^{+}.
\end{eqnarray}

We refer to this interaction channel as ``Channel (i)". Another prominant channel involves the interaction of an electron antineutrino with $^{12}\text{C}$, resulting in the production of a positron and $^{12}\text{B}$,
\begin{equation}
    \bar{\nu}_e + ^{12}\text{C} \rightarrow e^{+} + ^{12}B .
\end{equation}

This interaction is named as ``Channel (ii)". The third significant channel involves the interaction of an electron neutrino with $^{12}\text{C}$, resulting in the production of $^{12}\text{N}$ and an electron
\begin{equation}
    \nu_e + ^{12}\text{C} \rightarrow e^{-} + ^{12}N.
\end{equation}

This channel we term as ``Channel (iii)". Lastly, there is the elastic scattering interaction between an electron neutrino and an electron in the detector:
\begin{eqnarray}
    \nu_e + e^- \rightarrow \nu_e + e^- .
\end{eqnarray}
This channel is referred as ``Channel (iv)".

In addition to these four channels, we also consider neutral current (NC) interactions between various flavor neutrinos and $^{12}\text{C}$. The importance of including neutral current interactions lies in the fact that, in the presence of a sterile neutrino, the NC event rates differ, as some $\nu_e (\bar{\nu}_e)$ can convert to $\nu_s (\bar{\nu}_s)$ with a non-zero probability. This conversion results  a difference in event numbers between the active-active neutrino case and the active-sterile neutrino scenario.

Our study categorizes interaction channels separately for theoretical clarity, though NO$\nu$A, as a liquid scintillator detector, may not effectively distinguish charged-current interactions based on the final-state electron alone. Each interaction has distinct cross-sections and kinematic signatures—e.g., $\nu_e$-e scattering has a lower cross-section than IBD but provides directional information, while CC interactions on carbon involve nuclear de-excitation signatures. Due to NO$\nu$A's low spatial resolution, tracking individual particles is challenging, but the relative contributions of different interactions can still aid statistical analyses of the neutrino flux. While NO$\nu$A does not explicitly discriminate all channels in real-time, our separate analysis helps to evaluate theoretical models. Experimentally, detection is dominated by a combined CC signal, while NC interactions remain crucial for distinguishing active from sterile neutrinos, as they are independent of the charged-lepton final state.

\begin{figure}
    \centering
    \includegraphics[width=0.5\linewidth]{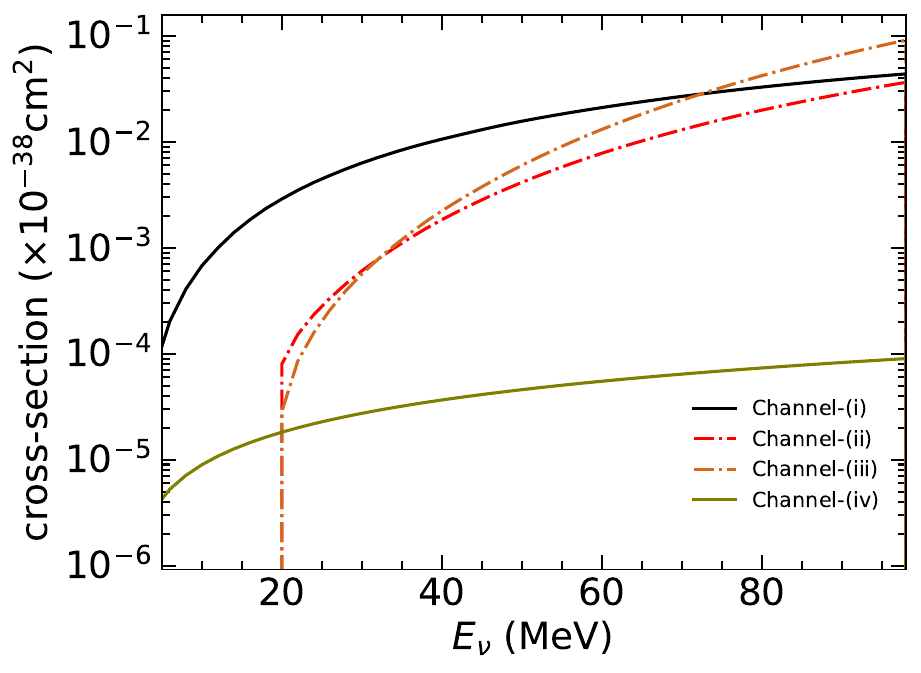}
    \caption{Cross section for different channels of NO$\nu$A detector.}
    \label{cross-section}
\end{figure}

\section{Event rates}
\label{event-rate-section}
In this section, we present the event rate distribution as a function of neutrino energy for the active-active scenario, and the active-sterile framework. Table \ref{appen-1} in Appendix \ref{app-a} displays the event rates of all the channels for both normal and inverted orderings in both the scenarios.
\begin{figure}
\includegraphics[scale=0.5]{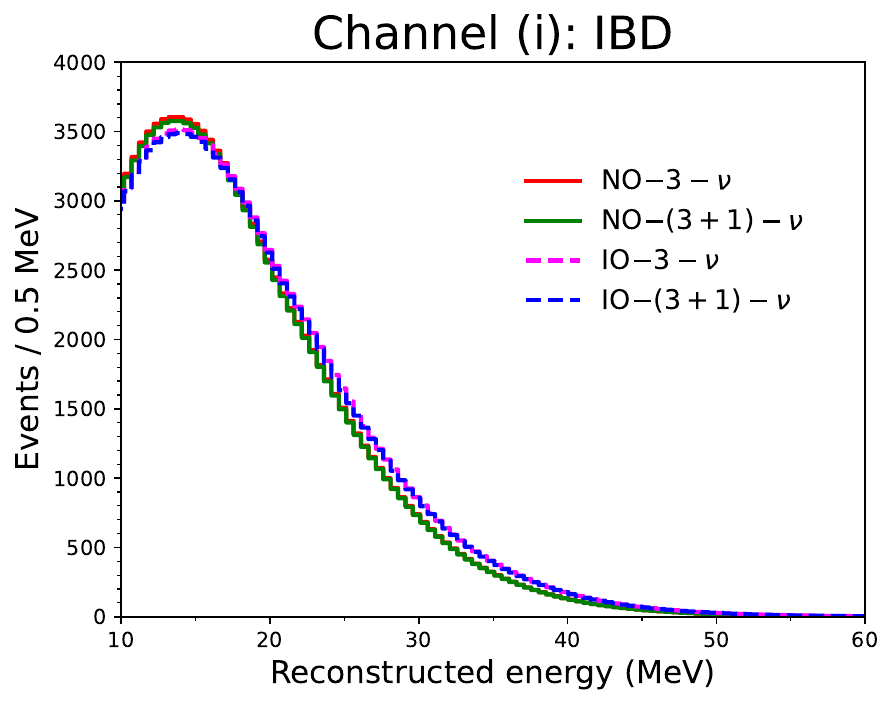}
\includegraphics[scale=0.5]{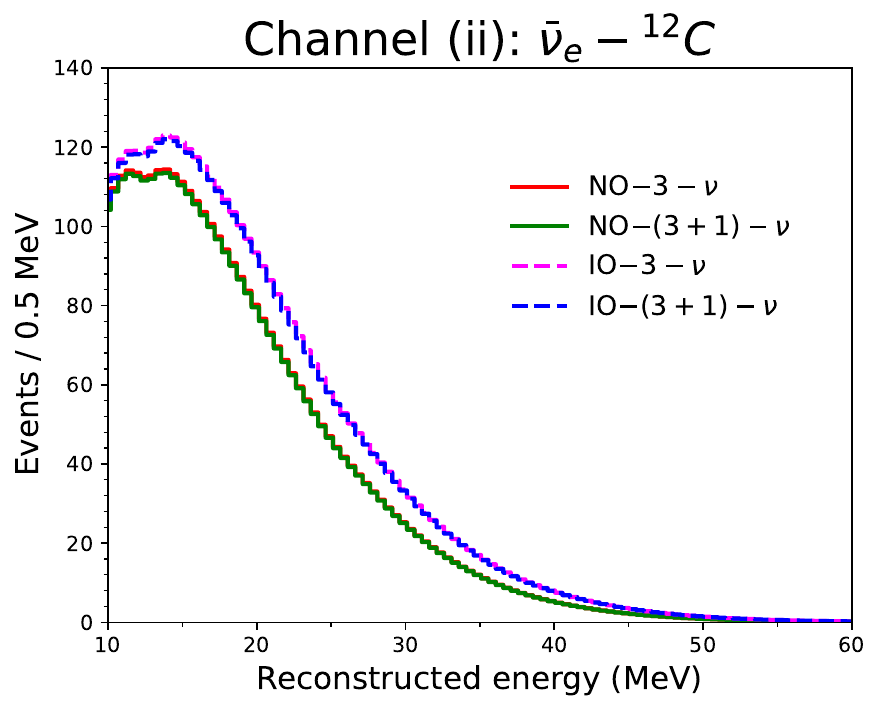}\label{eve-2}\\
\includegraphics[scale=0.5]{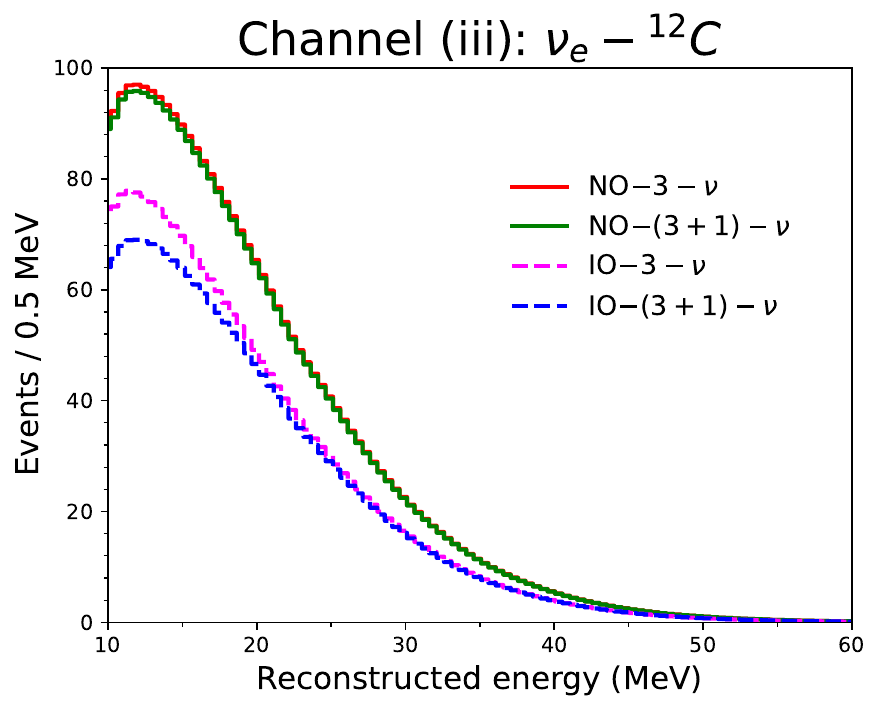}
\includegraphics[scale=0.5]{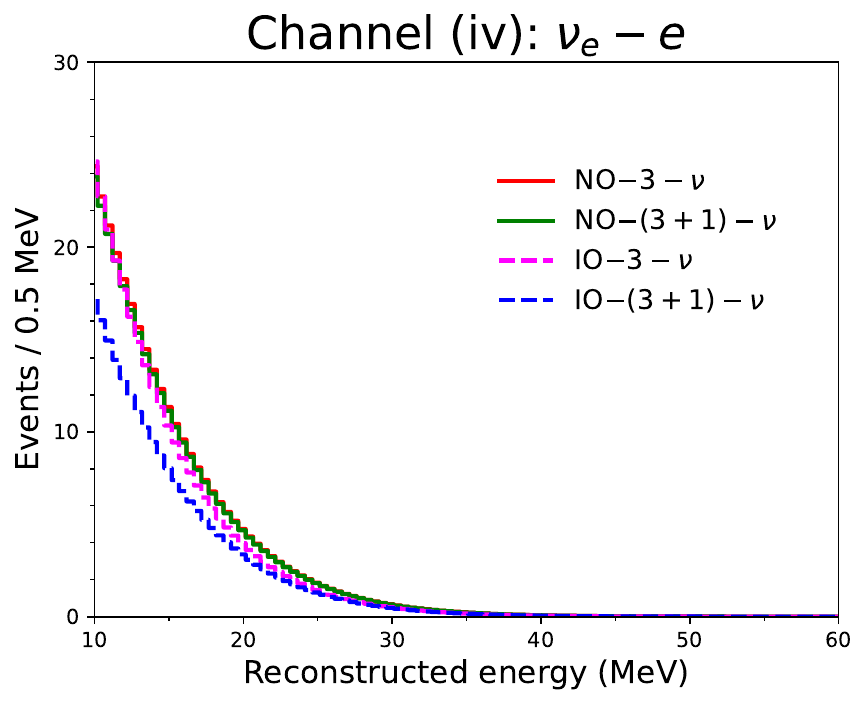}\\
    \includegraphics[scale=0.5]{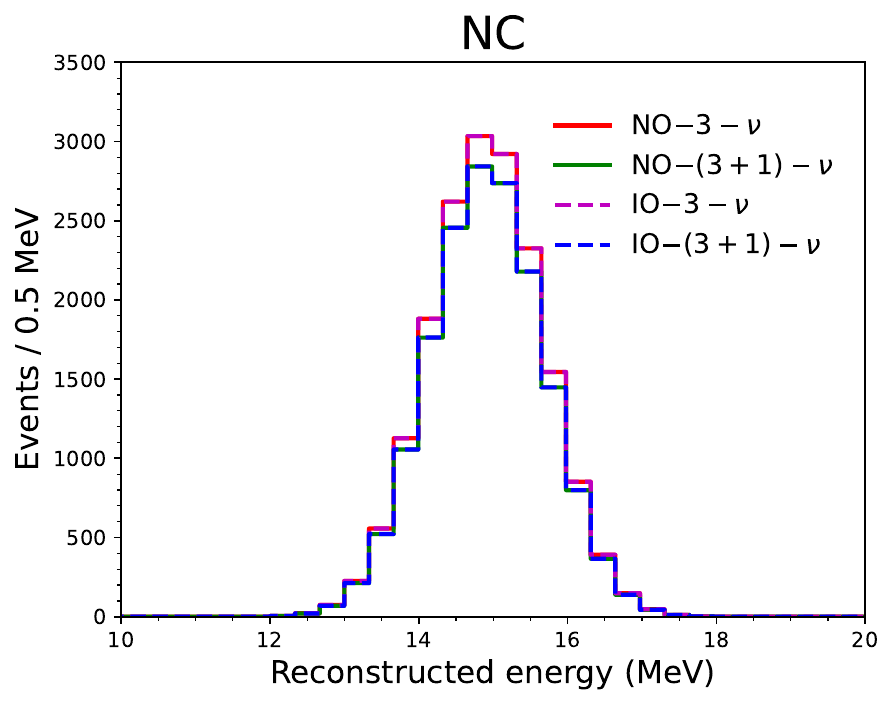}
    \caption{Event rate in active-active and active-sterile frameworks for five different channels for  supernova at a distance of 1 kpc. Color codes are given in the legend. NO (IO) represents the normal (inverted) ordering. }
    \label{f-2}
\end{figure}

To better understand the event rate behavior across different channels as a function of neutrino energy, we present Figure \ref{cross-section}, where the black, red dot-dashed, brown dot-dashed, and green curves correspond to the cross-sections for Channels (i), (ii), (iii), and (iv), respectively. It illustrates the cross-section behavior, helping in the interpretation of event rates. Since event rates result from the convolution of flux, cross-section, and detector response, analyzing the cross-section separately helps reveal the trends observed in Figure \ref{f-2}. In this figure, the IBD cross-section is the largest, followed by Channels (iii), (ii), and (iv). The event rate spectrum for each channel is obtained by multiplying its cross-section with the corresponding fluence.
To maintain consistency with background rejection strategies discussed in Ref. \cite{Vasel:2021ymu}, we have imposed a minimum reconstructed energy cut of 10 MeV in all event rate calculations. This cut ensures that the analysis focuses on the energy range where the NO$\nu$A detector can reliably distinguish supernova neutrino events from background.

Now, in the first subsection, we present the event rate distribution as a function of neutrino energy for Channels (i), (ii), (iii), and (iv) in the active-active scenario. After that, second subsection covers the event rate distribution for the channels mentioned above in the active-sterile scenario. In the next subsection, we also provide a detailed discussion of the neutral current channel to offer a comprehensive view. As the NO$\nu$A experiment detects all events collectively, the final row of Table \ref{appen-1} provides the total event rates for both normal and inverted mass orderings within the active-active and active-sterile scenarios.

\subsection{Active-active Neutrino Framework}

Table \ref{appen-1} shows the event numbers of the four different channels with NC event rates, assuming the supernova is located at a distance of 1 kpc. The first two rows for each channel represent the total event numbers for the active-active scenario, with both NO and IO conditions. From the Table, we observe that the total event number for Channel (i) is significantly higher than for Channel (ii), while the event numbers for Channel (iii) are relatively close to those of Channel (ii). In comparison, Channel (iv) has a much lower event count than Channel (iii) for both NO and IO conditions.
Similarly, the red and magenta dashed curves in each panel of Fig. \ref{f-2} visually illustrate the event rate spectra. From both Table \ref{appen-1} and Fig. \ref{f-2}, we observe that for Channel (i), there is a very small difference between the event rates for NO and IO. In Channel (ii), a significant difference emerges in the event rate spectrum after 10 MeV neutrino energy, with the IO event rates being higher than the NO rates. For Channel (iii), the event rates are higher for NO compared to IO. In contrast, for Channel (iv), the NO and IO event rates are nearly overlapping.

To understand the characteristics of the event rate spectrum, we focus on the fluence spectrum shown in Fig. \ref{flux}. For the event rate spectrum of Channel (i), we examine the fluence of $\bar{\nu}_e$ in Fig. \ref{flux}. It is observed that for $\bar{\nu}_e$, the fluences for NO and IO overlap, which explains the event rate spectrum behavior for Channel (i). Similarly, for Channel (ii), we again focus on the $\bar{\nu}_e$ fluence. In this case, we see that the event rate for the inverted ordering is higher than that for the normal ordering. This is due to the interplay between the fluence and the cross-section of Channel (ii). The event rate is proportional to the product of the fluence and the cross-section for the particular channel. Turning to Channel (iii), this channel involves the interaction of an electron neutrino with $^{12}\text{C}$. In this case, we focus on the $\nu_e$ fluence, and here the IO fluence is higher than the NO fluence. However, when the cross-section is multiplied by the fluence, the event rate for NO becomes greater than that for IO. Channel (iv) corresponds to the elastic scattering of an electron with an electron neutrino. It is interesting to note that the event rate behavior for Channel (iv) differs from the other channels. This difference is attributed to the smearing effect in the detector. If the smearing effects were removed, the event rate behavior would resemble that of the other channels.

\subsection{Active-sterile neutrino framework}

In this subsection, we present the event rates in the active-sterile framework. The last two rows of each channel in Table \ref{appen-1} show the event numbers for the active-sterile scenario. 
It is evident that in the presence of a sterile neutrino, the detected event rate decreases, as some of the active neutrinos oscillate into the undetected sterile neutrinos. Table \ref{appen-1} indicates that, for both NO and IO conditions, the event number in the active-sterile scenario is lower than that in the active-active scenario. 

The results from Table \ref{appen-1} are graphically represented in Fig. \ref{f-2}. In each panel of Fig. \ref{f-2}, the green and blue-dashed curves represent the event rates for the NO and IO active-sterile neutrino frameworks, respectively. The explanations for the curves in each channel are similar to those provided for the active-active framework.

\subsection{NC channels}

Now, we discuss the neutral current event rates for both the active-active and active-sterile scenarios. Neutral current interactions apply to all neutrino flavors. In our calculation, we have considered the interactions of all neutrino flavors with $^{12}\text{C}$, as the primary material in the NO$\nu$A detector is Carbon-12.
The last row of Table \ref{appen-1} presents the total number of neutral current (NC) events, obtained by summing all six types of NC interactions, at a distance of 1 kpc for both normal and inverted mass orderings. The Table includes scenarios for both the active-active neutrino case and the active-sterile neutrino framework. The differences in event rates between these scenarios highlight the need to explore the impact of sterile neutrinos on supernova neutrino oscillations.

\section{Results}
\label{results}
In this section we show the main results of our work for both the neutrino frameworks. We divide our results some important subsections while in each subsection, we illustrate the results drawn from active-active scenario  first, and then  talk about the results from the active-sterile framework. All results have been derived from neutrino event rates above 10 MeV, taking into account the applied background rejection techniques.

\subsection{Mass ordering sensitivity}

The sensitivity to the neutrino mass ordering refers to an experiment's ability to distinguish between normal and inverted mass ordering in the neutrino spectrum. Supernovae, being prolific sources of neutrinos, provide an excellent opportunity to probe the mass ordering. This can be achieved through currently operating long-baseline neutrino experiments, such as NO$\nu$A. In this subsection, we discuss how the mass ordering sensitivity varies as a function of the supernova distance (measured in kpc).

\begin{figure}
    \centering
    \includegraphics[scale=0.5]{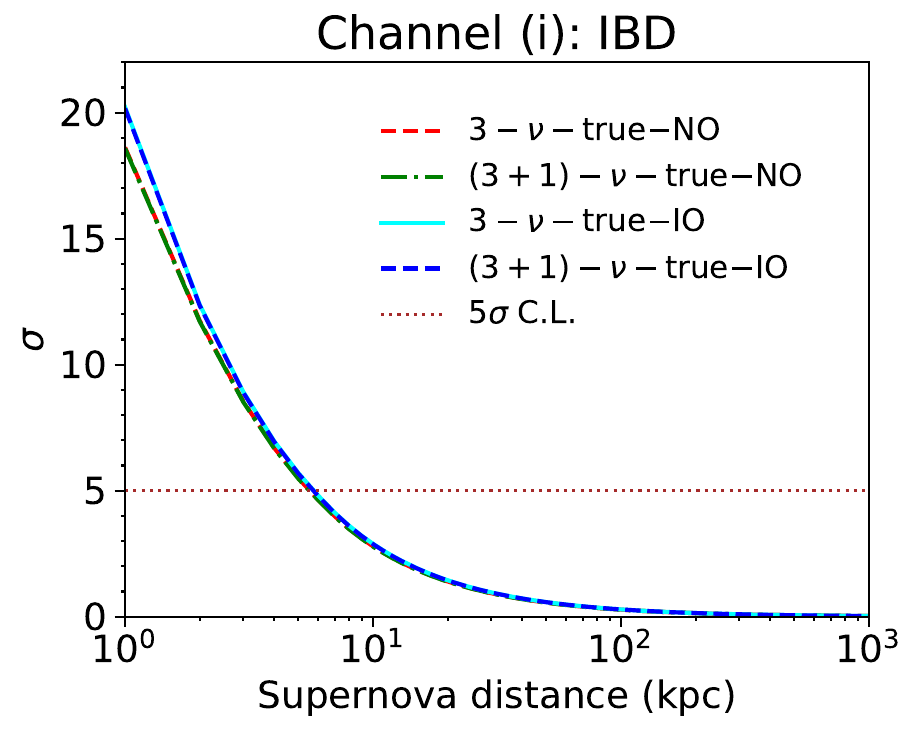}
    \includegraphics[scale=0.5]{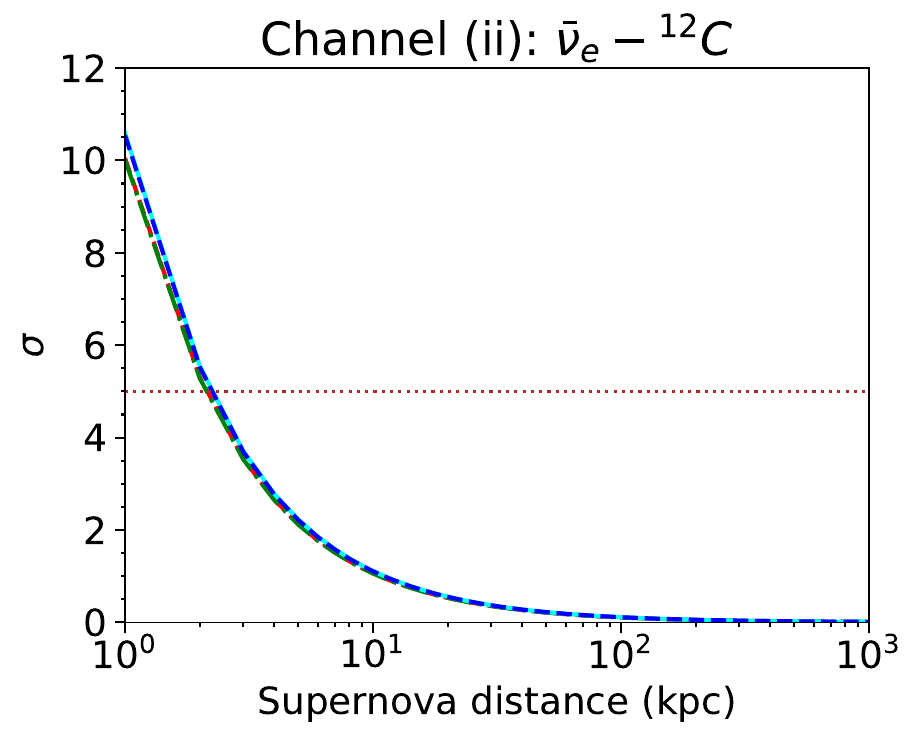}
    \\
    \includegraphics[scale=0.5]{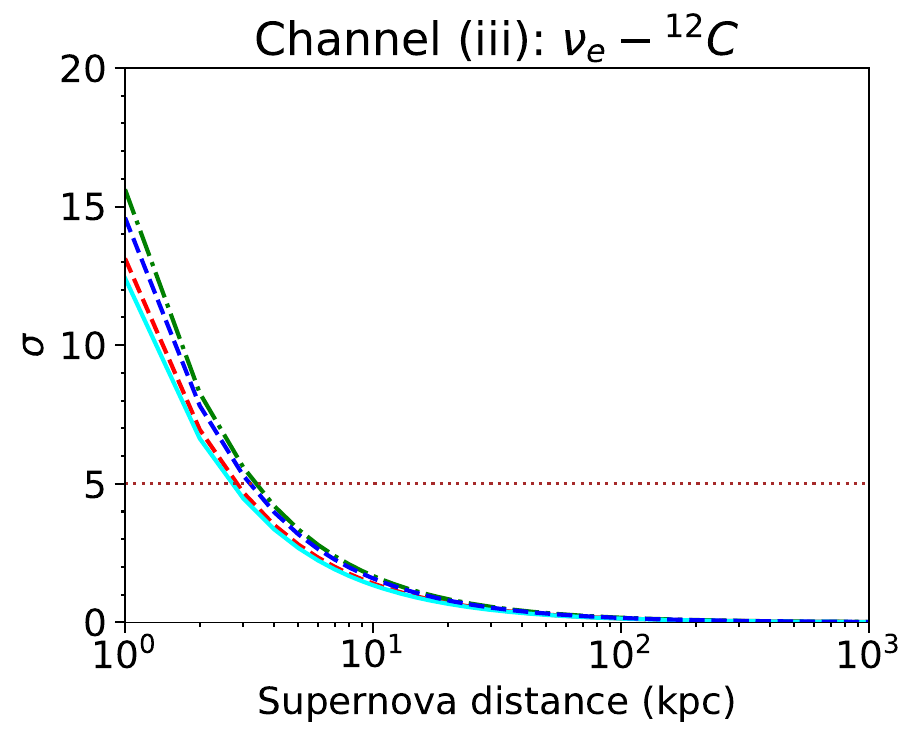}
    \includegraphics[scale=0.5]{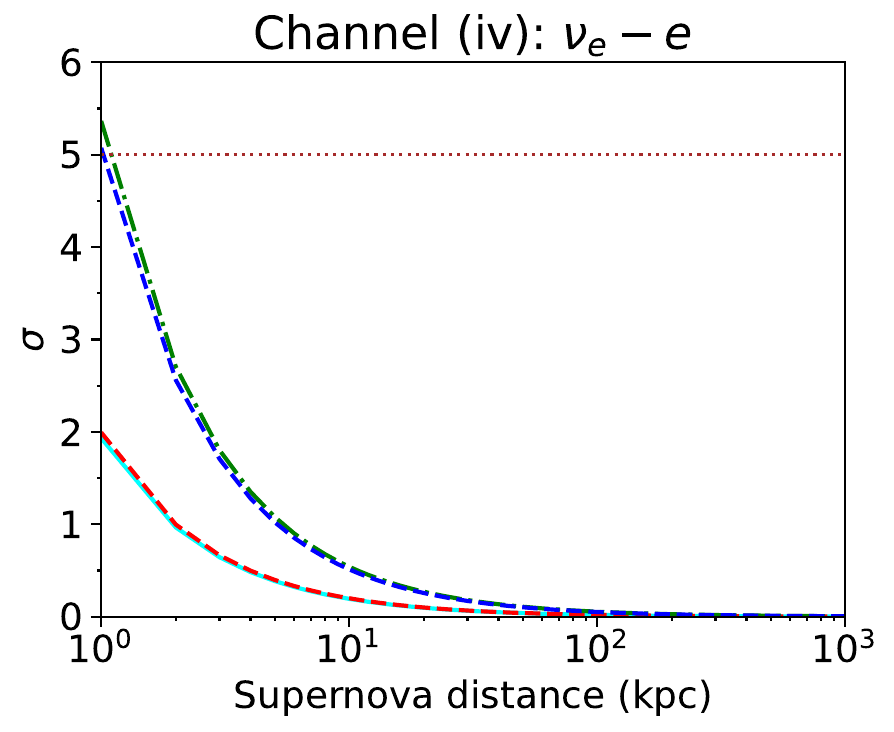} 
    \includegraphics[scale=0.5]{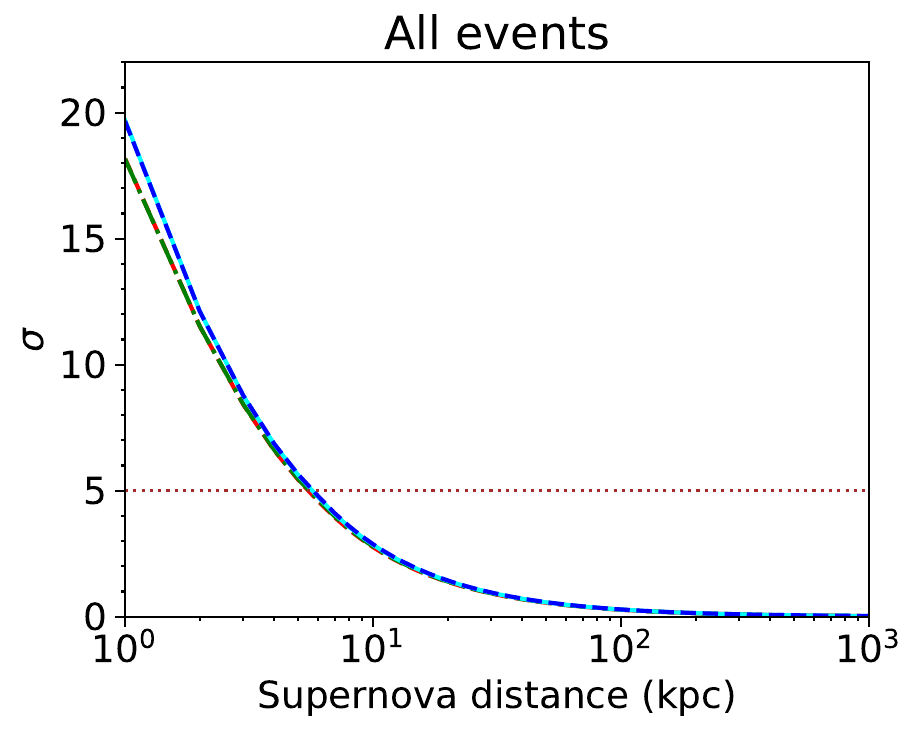}
    \caption{Mass ordering sensitivity ($\sigma = \sqrt{\Delta \chi^2}$) as a function of supernova distance (in kpc). Color codes are given in the legends. }
    \label{f-3}
\end{figure}

The key findings are presented in Fig. \ref{f-3}. Each panel illustrates the mass ordering sensitivity as a function of the supernova distance for various detection channels. In our analysis, $\sigma$ corresponds to the statistical significance of the mass ordering sensitivity and is defined as:  
$\sigma = \sqrt{\Delta \chi^2}$
where $\Delta \chi^2$ represents the chi-squared difference relative to the minimum chi-squared value. We consider four primary channels, and the results for all four channels are shown in Fig. \ref{f-3} under four different scenarios. For the case labeled $``3-\nu-$true$-$NO (IO)", the analysis assumes a active-active scenario with the true spectrum having normal (inverted) ordering and the test spectrum as inverted (normal) ordering. Similarly, the condition $``(3+1)-\nu-$true$-$NO (IO)" corresponds to the active-sterile framework, where the true spectrum has a normal (inverted) ordering and the test spectrum with inverted (normal) ordering. 

In each panel, the red dashed (green dot-dashed) curve represents the mass ordering sensitivity for the active-active (active-sterile) scenario with  normal ordering in the true spectrum. Likewise, the cyan solid (blue dashed) curve shows the results for the active-active (active-sterile) scenario with  inverted ordering in the true spectrum. The brown dotted line indicates the $5\sigma$ confidence level threshold for mass ordering sensitivity. 
Fig. \ref{f-2} presents the event rates for each interaction channel under different scenarios: standard $3-\nu$, and $(3+1)-\nu$ with both normal and inverted mass orderings. In contrast, Fig. \ref{f-3} depicts the sensitivity to the mass ordering for these channels, considering either normal or inverted ordering as the true spectrum, within both $3-\nu$ and $(3+1)-\nu$ frameworks. The color schemes used in Fig. \ref{f-2} and Fig. \ref{f-3} are intentionally kept distinct, reflecting their different physical meanings and the fact that they are not directly related for comparison.
From Fig. \ref{f-3}, we observe that sensitivity is generally higher when the true spectrum assumes as inverted ordering compared to normal ordering. Examining the individual channels we found that for channels (i) and  (ii), both active-active and active-sterile scenarios produce nearly identical results. The results indicate that using only the IBD channel, NO$\nu$A can discriminate the correct mass ordering from the incorrect one at the $5\sigma$ confidence level for a supernova located approximately 5 kpc away. However, for channels (iii) and (iv), a clear distinction emerges, with the active-sterile framework showing a higher sensitivity than the active-active scenario. This behavior can be explained using Table \ref{appen-1}, which highlights a significant discrepancy in event numbers between normal and inverted hierarchies for the active-sterile framework. In contrast, the active-active scenario exhibits minimal differences in event numbers between these orderings.

For the NC channel, 
as the event rate for NO and IO condition of active-active scenario is same in number, there is no mass ordering sensitivity from this channel.
However, in the active-sterile framework, a non-zero difference between NO and IO event rates introduces some mass ordering sensitivity. Since the difference in event numbers for NO and IO in the active-sterile framework is small for the NO$\nu$A detector, this setup does not offer significant mass ordering sensitivity. Notably, future long-baseline experiments with much larger far detector volumes are expected to show a significant difference in event numbers between the active-active and active-sterile scenarios, as well as between NO and IO event rates in the active-sterile framework, leading to significant mass ordering sensitivity from the NC channel.
Thus, the NC channel emerges as a valuable tool for investigating the possible existence of sterile neutrinos in nature.

The bottom panel of Fig. \ref{f-3} depicts the mass ordering sensitivity of supernova neutrinos when all event channels are combined. It follows the same color coding as the earlier panels in Fig. \ref{f-3}. The results closely resemble those of channel (i), demonstrating that the NO$\nu$A experiment can distinguish between normal and inverted mass orderings with a 5$\sigma$ confidence level if the supernova occurs at a distance of 5 kpc from Earth.

\subsection{Effect of systematics}
\begin{figure}
\centering
\includegraphics[scale=0.5]{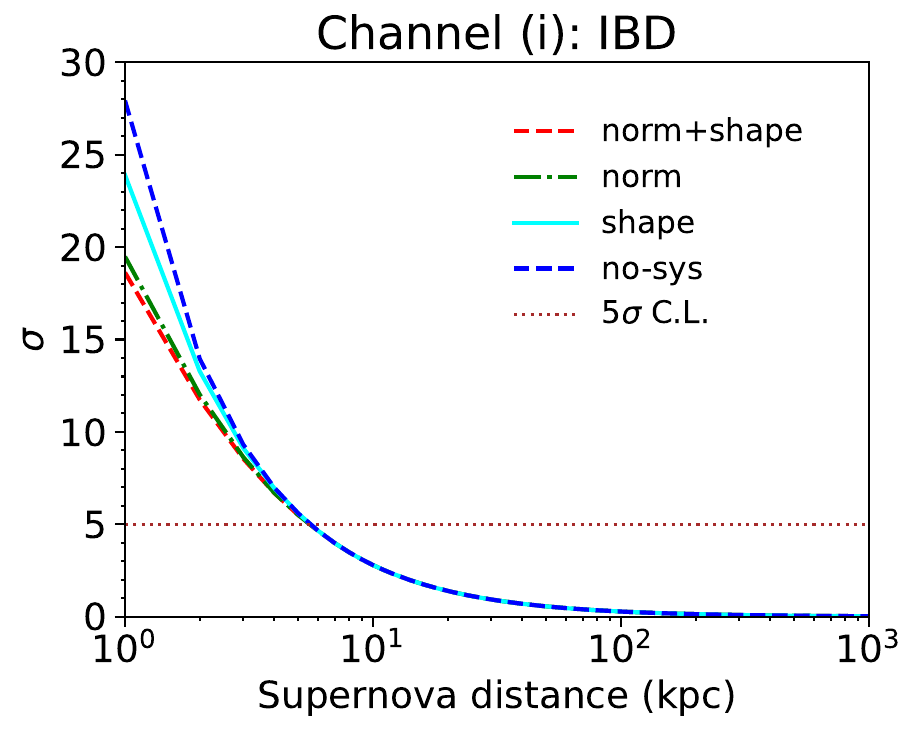}
\includegraphics[scale=0.5]{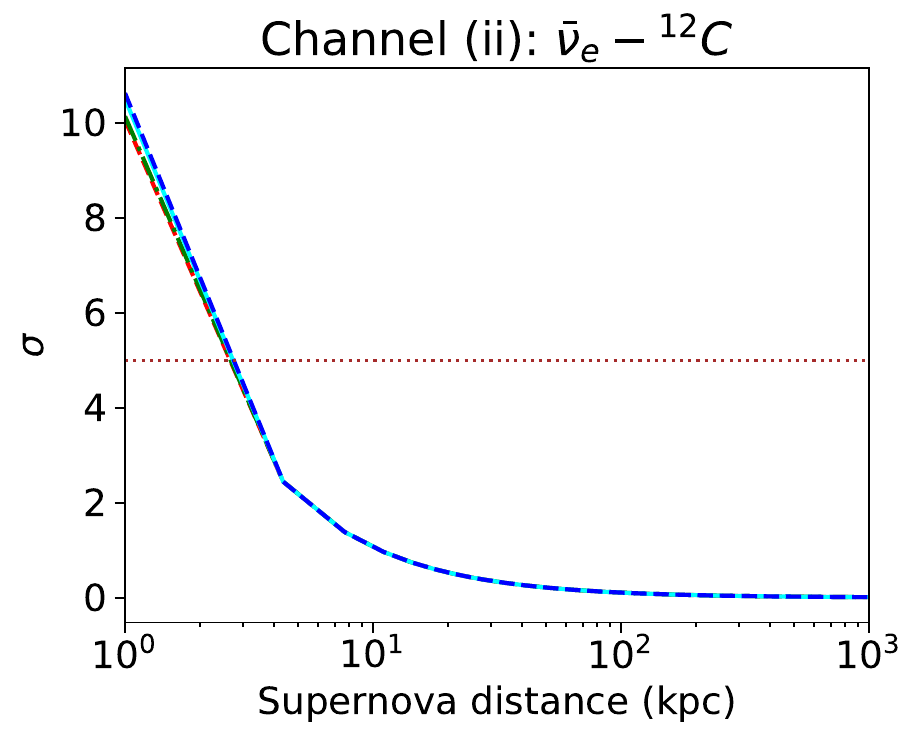}\\
\includegraphics[scale=0.5]{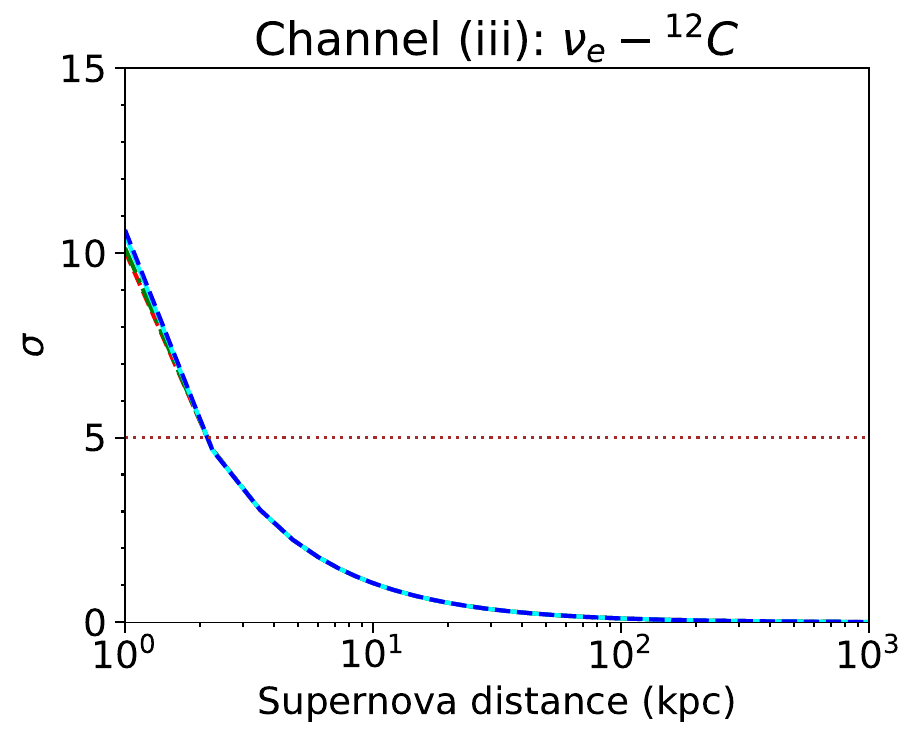}
\includegraphics[scale=0.5]{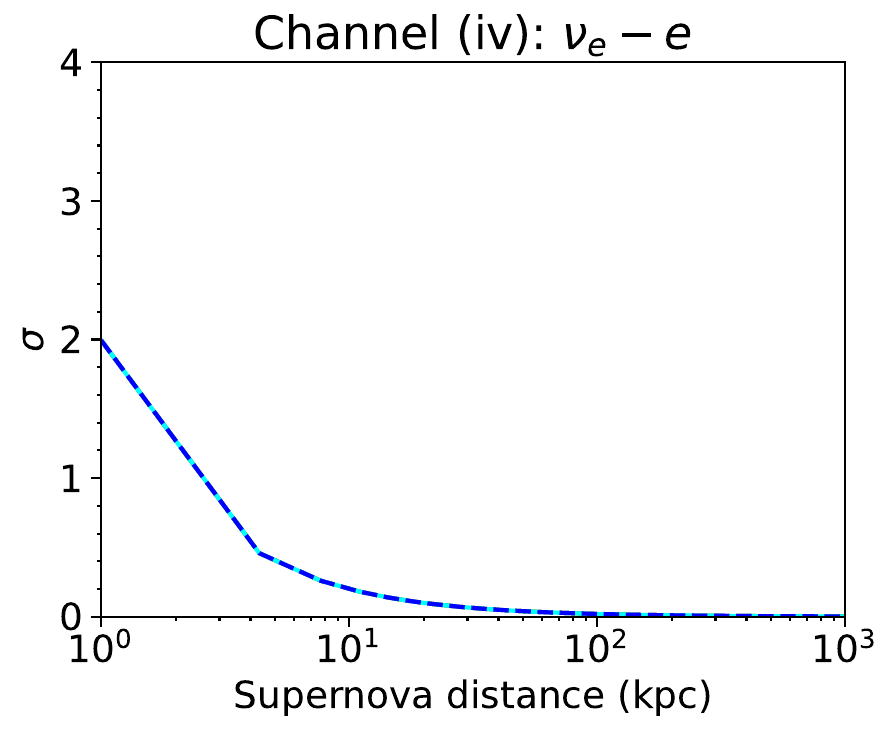}
    \caption{Mass ordering sensitivity as a function of supernova distance (in kpc) for all four channels in different systematics uncertainty conditions; ``norm" (``shape") stands for normalization (energy calibration) error. This figure is for active-active framework, however similar nature is for active-sterile scenario. Color codes are given in the legend.}
    \label{sys-ordering}
\end{figure}

In this subsection, we examine the impact of systematic uncertainties on mass ordering sensitivity as a function of supernova distance. These uncertainties arise from various sources, including supernova flux measurements, cross-section measurements, and the incoming neutrino direction. To analyze these effects, we present Fig. \ref{sys-ordering}, where each channel is shown with curves corresponding to different combinations of systematic errors. 
The Figure shows the results for active-active scenario only. Similar effect can be seen in active-sterile framework as well.  

\begin{figure}
\includegraphics[scale=0.5]{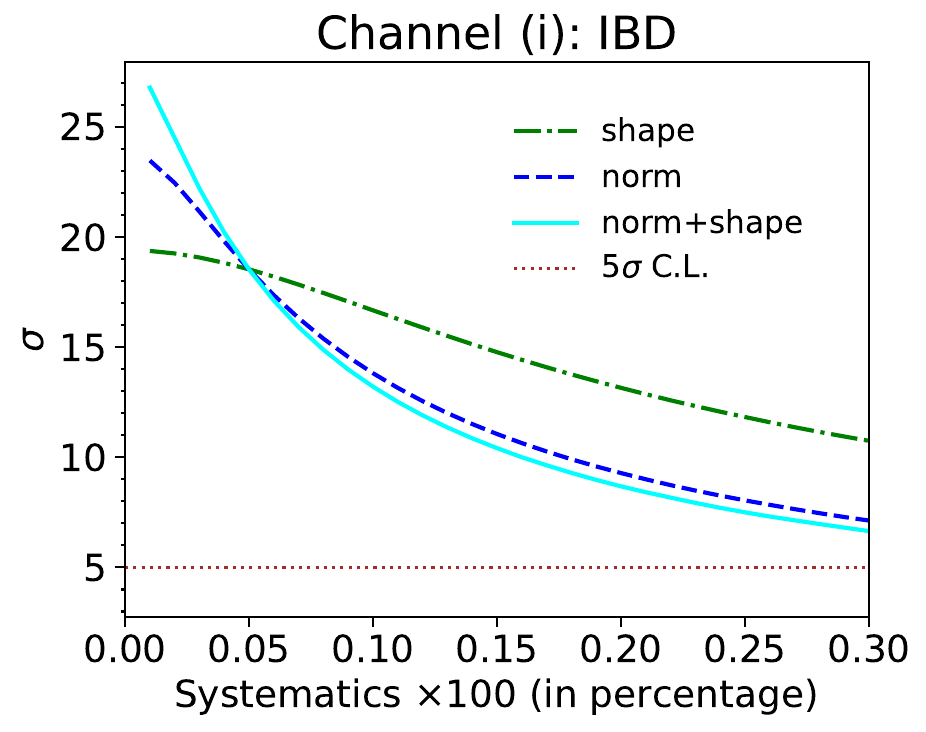}
\includegraphics[scale=0.5]{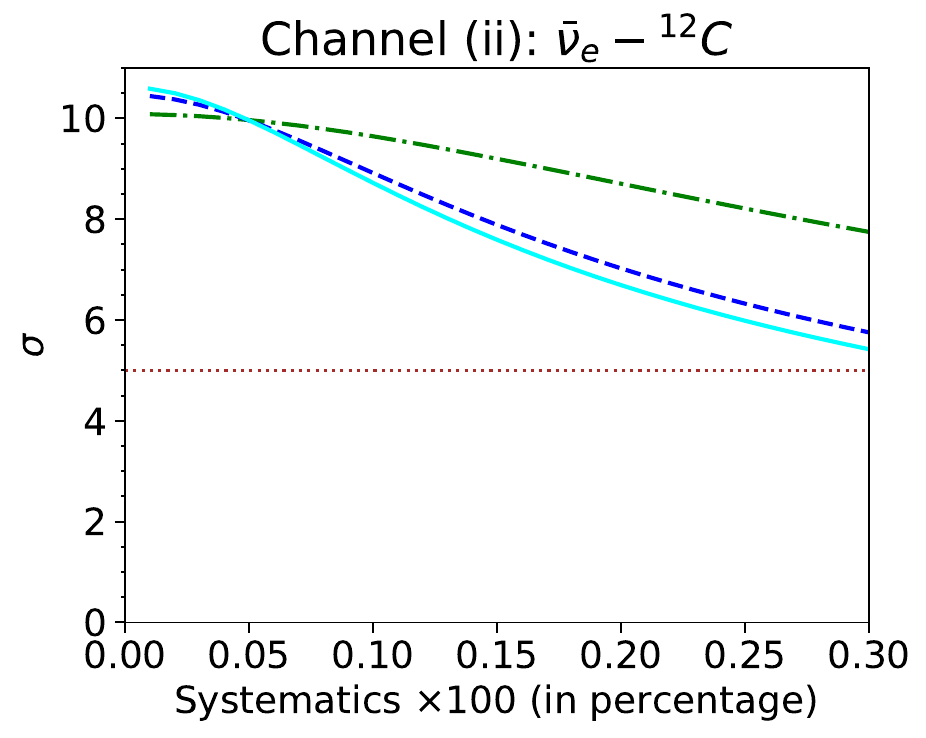}
\\
\includegraphics[scale=0.5]{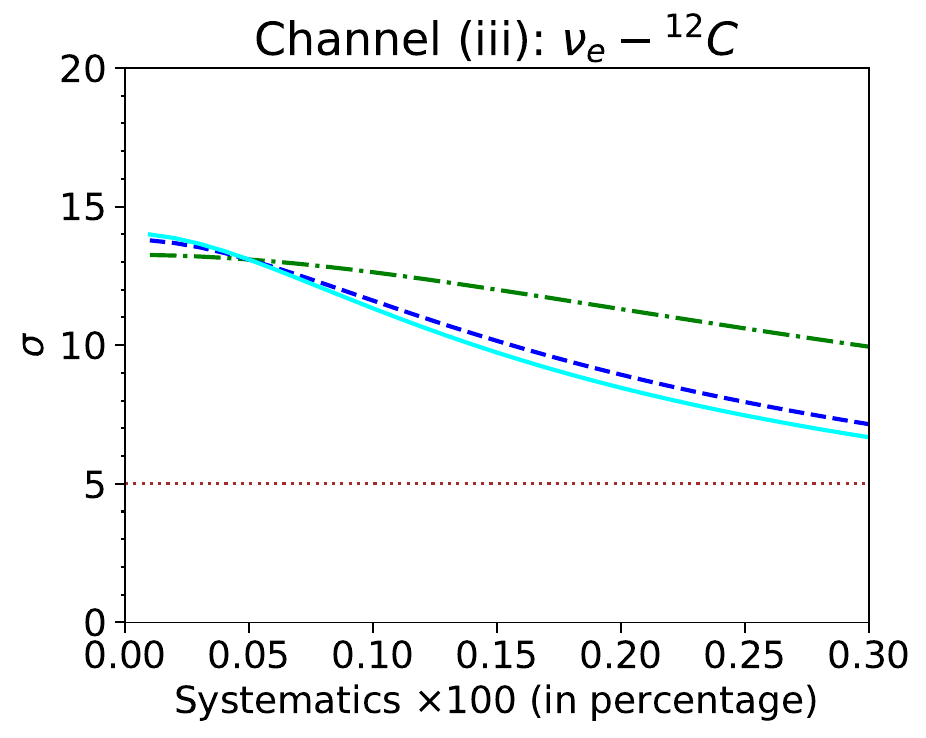}
\includegraphics[scale=0.5]{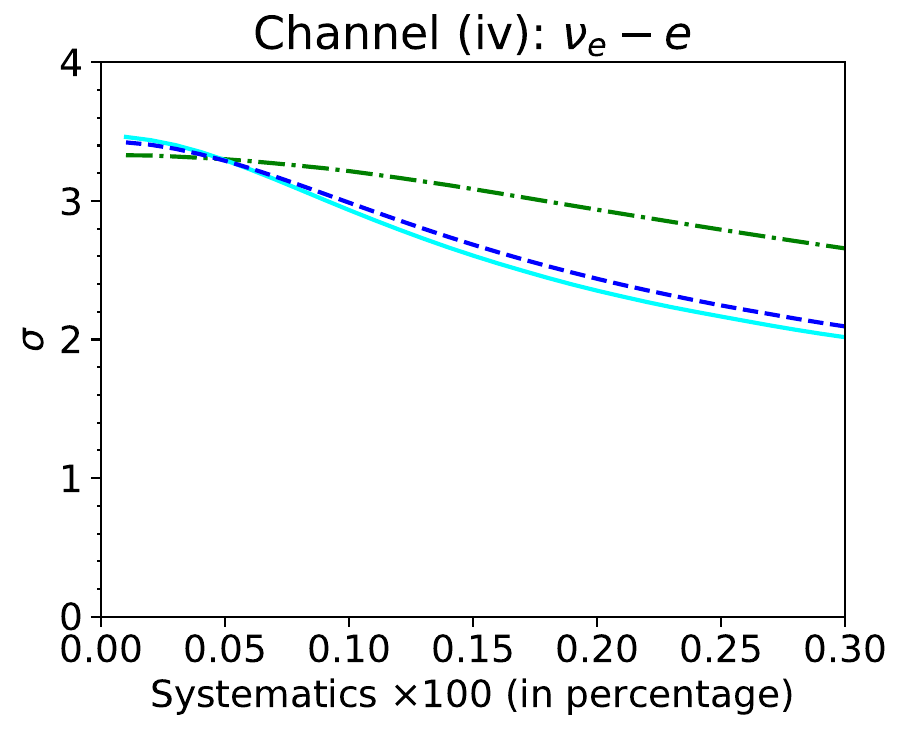}
    \caption{Mass ordering sensitivity as a function of systematic uncertainty (in percentage) at supernova distance of 1 kpc for different conditions. This figure is for active-active framework. However, similar nature is for active-sterile scenario. Color codes are in the legend. }
    \label{sys-vary}
\end{figure}

In our analysis, we consider two types of errors: normalization error and energy calibration error. In each panel, ``norm" represents normalization uncertainties, while ``shape" corresponds to energy calibration uncertainties. For the cyan (green dot-dashed) curve, we set $p_1$ ($p_2$) to zero in Eq. \ref{chi-sys}. The red dashed curve includes contributions from both $p_1$ and $p_2$ in the $\chi^2$ calculation. The figure illustrates that when both uncertainties are considered, the mass ordering sensitivity is lower compared to cases where only one uncertainty is present. The blue dashed curve, representing the scenario with no systematic errors, exhibits the highest sensitivity. 

Interestingly, normalization errors have a more pronounced impact on ordering sensitivity compared to energy calibration errors, as the deterioration in sensitivity is greater when only normalization errors are included. This trend is consistent across all channels and for both active-active and active-sterile frameworks.

To provide a clearer understanding of how systematic errors influence ordering sensitivity, we present Fig. \ref{sys-vary}. In each panel of this figure, the green dot-dashed curve corresponds to scenarios with only energy calibration errors, while the blue dashed curve includes only normalization errors. The cyan solid curve accounts for both errors simultaneously. The figure depicts mass ordering sensitivity as a function of systematic uncertainty, specifically the coefficient of a and b in eq. \ref{chi-sys},  which is varied from $0\%$ to $30\%$, for a supernova distance of 1 kpc.

As systematic uncertainty increases from $0\%$ to $30\%$, the sensitivity decreases from $25\sigma$ to $12\sigma$ for the primary IBD channel. Notably, all error types intersect at a systematic uncertainty of $5\%$.
For the NC channel, however, ordering sensitivity remains unaffected by changes in systematic error. 
This is because the mass ordering sensitivity for the active-sterile framework in the NC channel is inherently very small, making it insensitive to the type or magnitude of the systematic error. Therefore, we have not shown it in our result panel. 
In conclusion, we can say that, systematic uncertainty has a significant impact on the mass ordering sensitivity in every channel.
\begin{figure}
\includegraphics[scale=0.5]{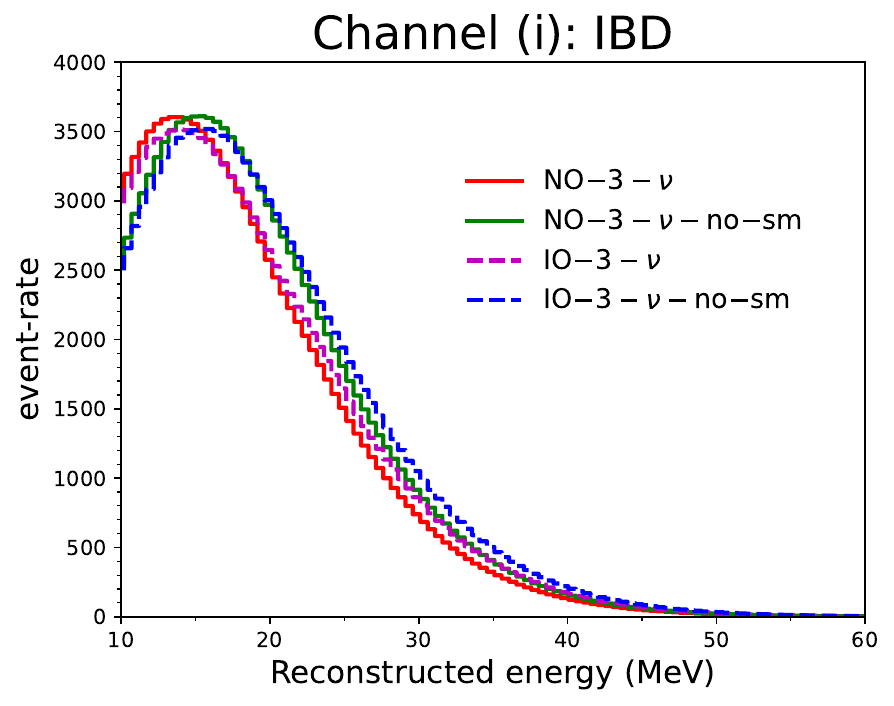}
\includegraphics[scale=0.5]{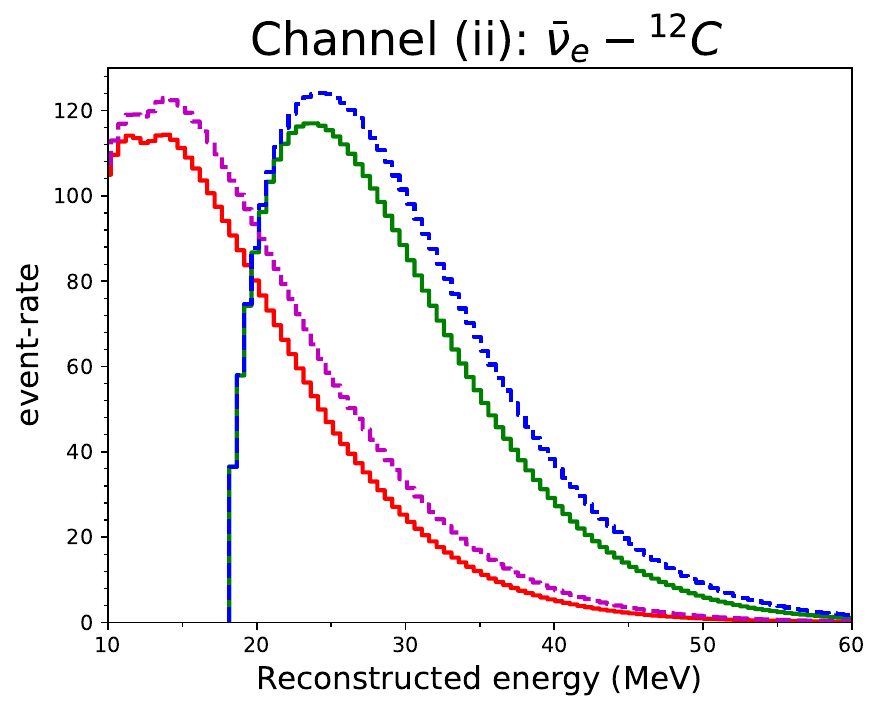}\\
\includegraphics[scale=0.5]{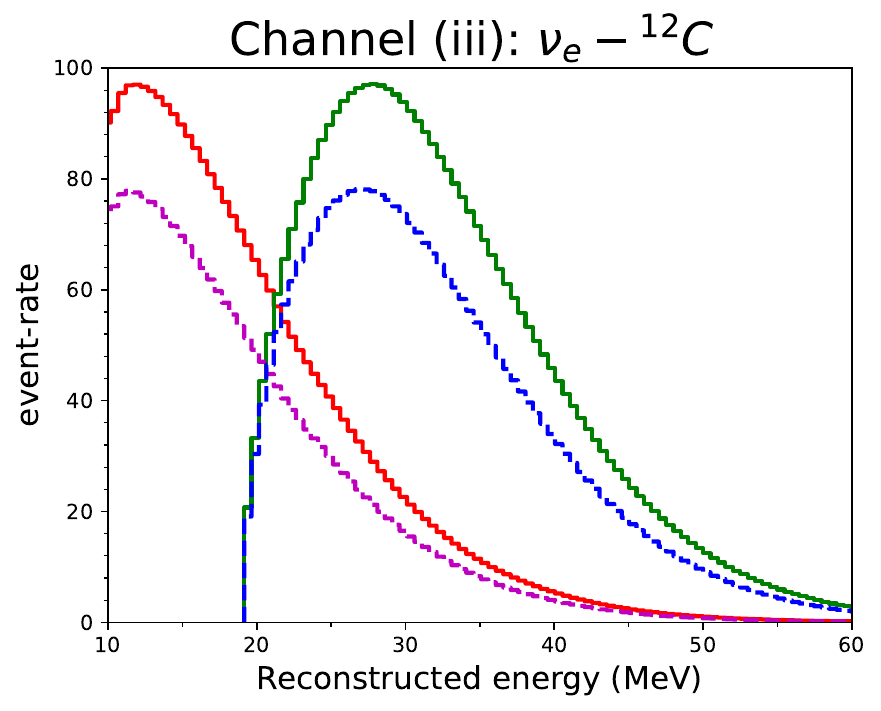}
\includegraphics[scale=0.5]{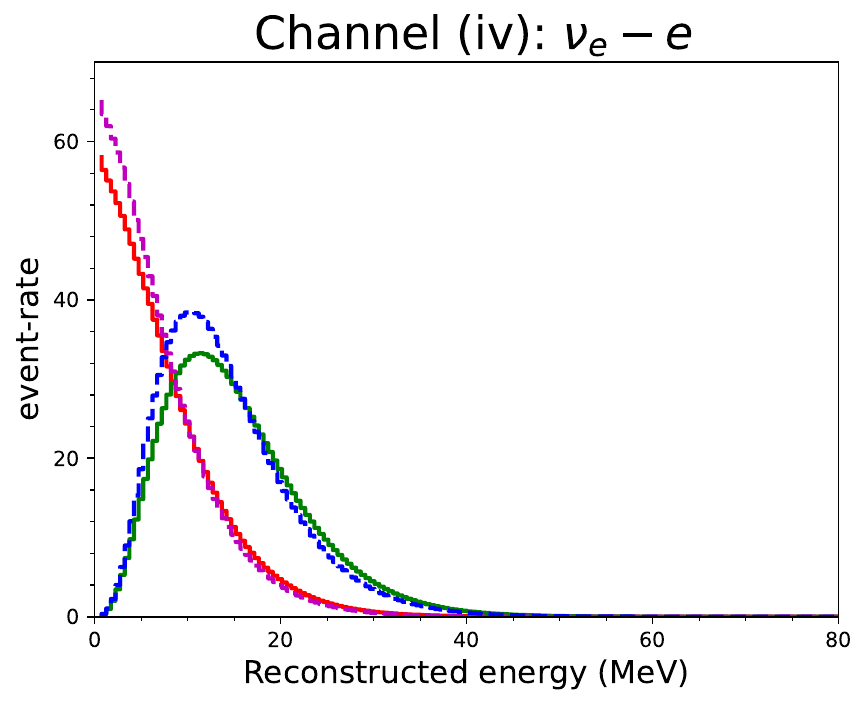}
\\
\includegraphics[scale=0.5]{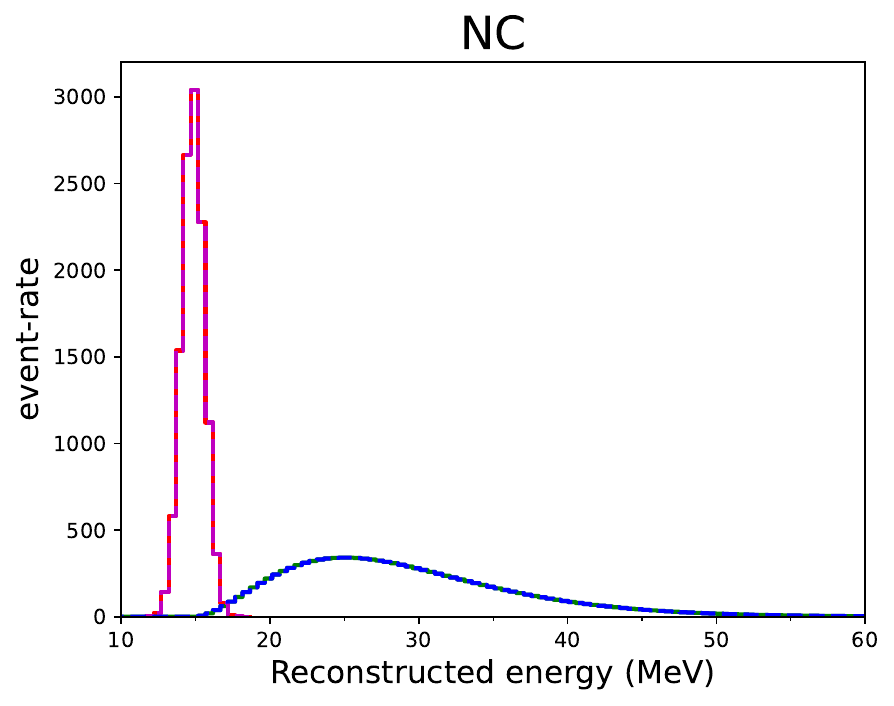}
    \caption{Event rate for active-active framework as a function of neutrino energy (MeV) for all the main channels and NC. All the plots are for supernova distance 1 kpc. Similar nature has been shown for active-sterile scenario also. Here sm [no-sm] refers to the terms with [without] smearing matrix. Color codes are given in the legend of each panel.  }
    \label{f-6}
\end{figure}

\subsection{Effect of smearing}
\begin{figure}
    \includegraphics[scale=0.5]{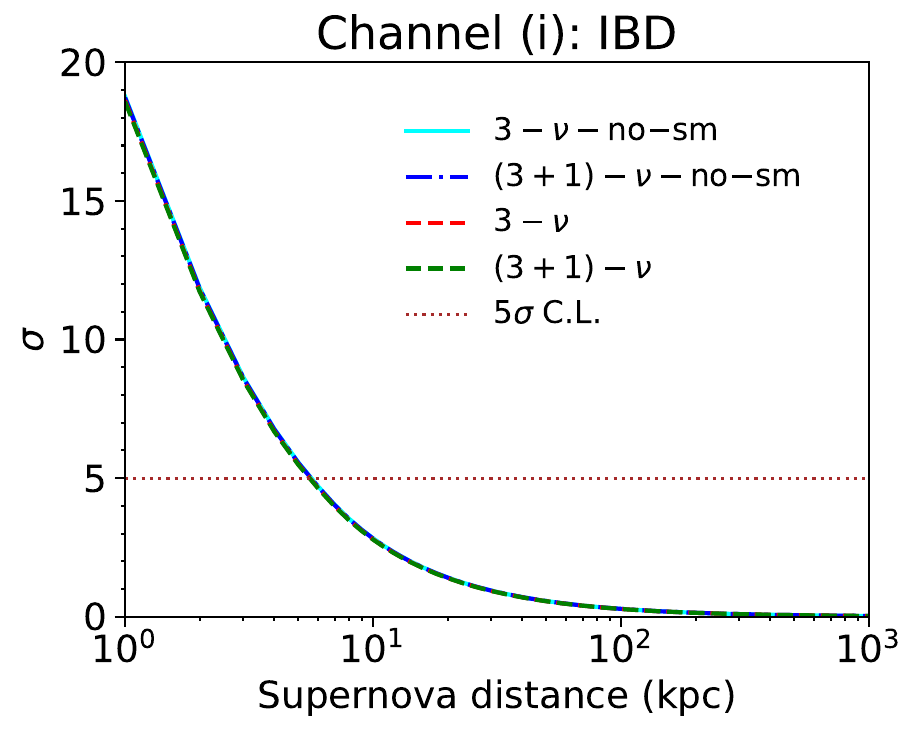}
    \includegraphics[scale=0.5]{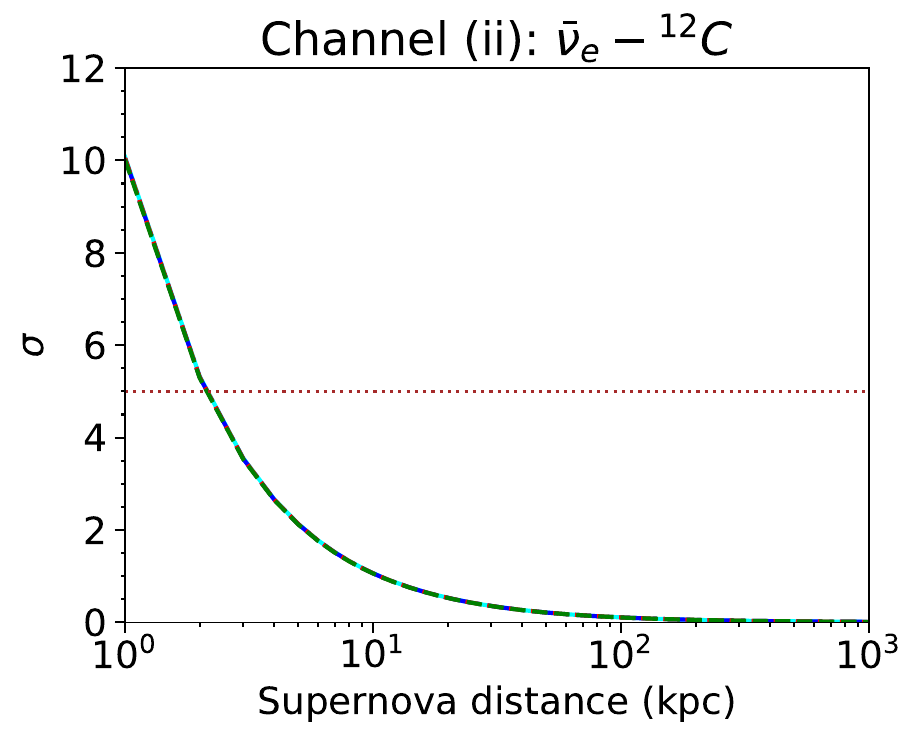}\\
    \includegraphics[scale=0.5]{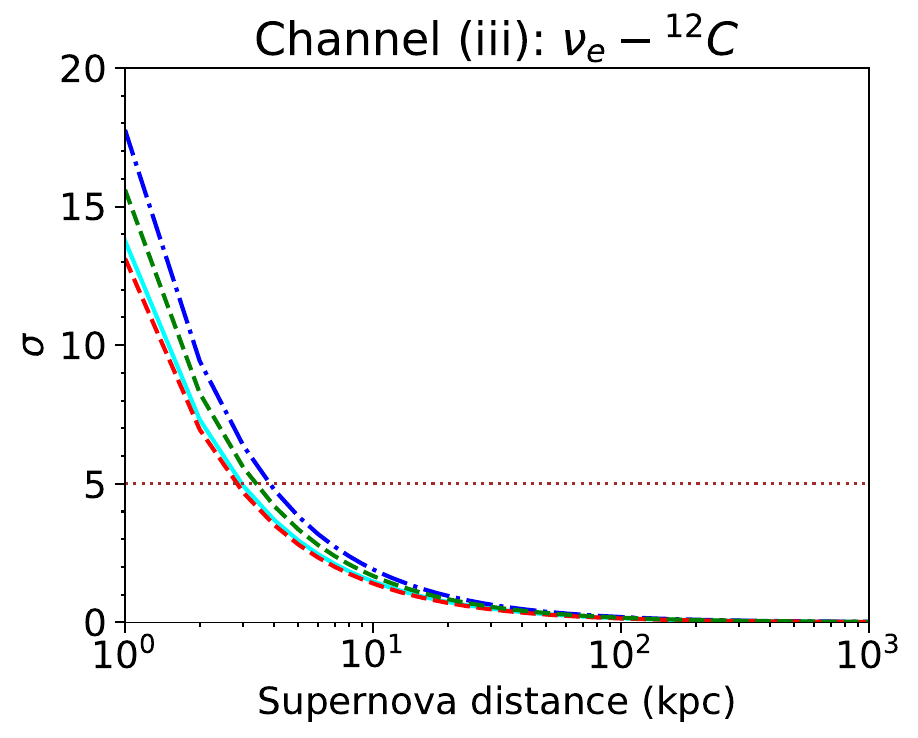}
    \includegraphics[scale=0.5]{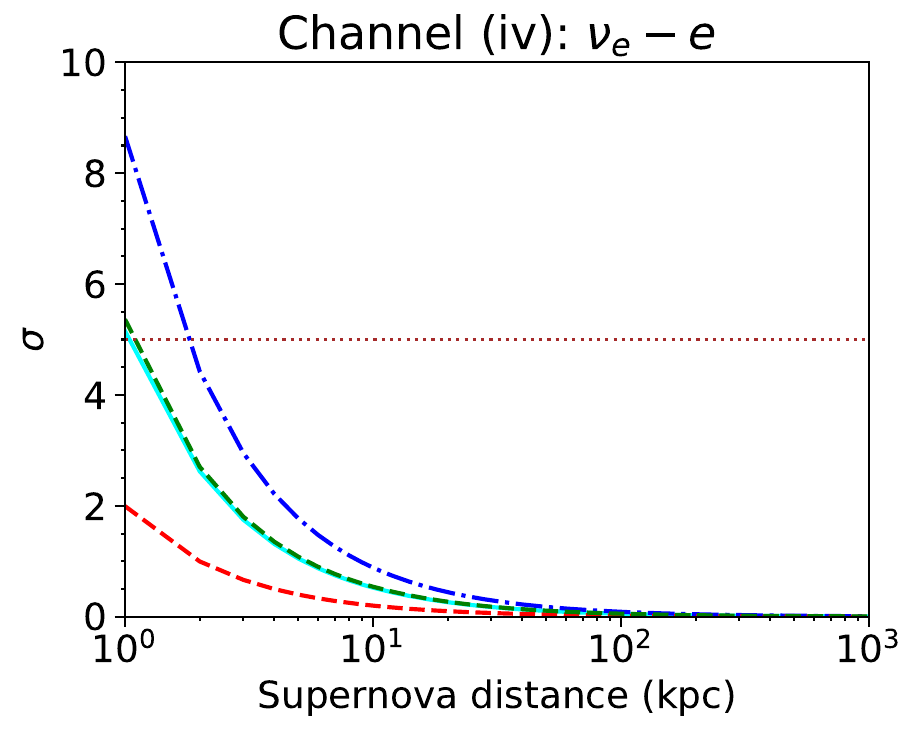}
    \caption{Mass ordering sensitivity as a function of supernova distance (in kpc) with [without] smearing matrix condition for all the channels. Here sm [no-sm] refers to the terms with [without] smearing matrix. Color codes are given in the legend of each panel.  }
    \label{f-7}
\end{figure}
In any long-baseline neutrino experiment, the 
energy of the neutrino is determined by measuring the energy of the outgoing leptons. This process inherently leads to a loss of information from the incoming neutrino, a phenomenon referred to as ``energy smearing." The inclusion of energy resolution in the analysis results in a further loss of information in the detector, thereby reducing the sensitivity to the neutrino mass ordering. 

In this subsection, we investigate the impact of energy smearing on mass ordering sensitivity for supernova neutrinos in both active-active and active-sterile frameworks. Fig. \ref{f-6} illustrates how energy smearing affects the event rate spectrum, comparing the cases with and without smearing. This figure represents the event rate spectra for the active-active framework. Similar effect can also be seen for active-sterile scenario as well. In each panel, the red (green) and magenta dashed (blue-dashed) curves correspond to the event rates for NO (IO) conditions with the presence (absence) of smearing. 
 
The discussion highlights that for channels (i), (ii), and (iii), the event rate spectra remain similar in shape but shift leftward due to energy smearing, as the smearing reduces the reconstructed energy of events.
The energy shift observed in Figure \ref{f-6} arises from the effect of detector energy resolution, which we model through energy smearing. In realistic detectors, the reconstructed neutrino energy is estimated based on the measured energy and angle of the outgoing lepton. However, this reconstruction is inherently limited by the detector’s finite resolution.
We account for this uncertainty using a Gaussian smearing function, which causes events to be reconstructed at energies slightly different from their true values. As a result, the reconstructed energy spectrum differs from the true energy distribution.
This smearing leads to a shift of spectral features toward lower energies, while the overall shape of the spectrum remains qualitatively similar. The energy resolution effect reduces the accuracy of energy reconstruction and consequently leads to a decrease in the expected sensitivity.

In contrast, for channel (iv), energy smearing modifies the shape of the event rate spectrum. Similarly, for the NC channel, the spectrum becomes more compact with energy smearing, whereas, without smearing, the spectrum is more widely spread.  

Next, Figure \ref{f-7} illustrates the impact of smearing on mass ordering sensitivity as a function of supernova distance. The results show that energy smearing reduces the sensitivity across all the channels. 
Improved energy resolution enhances sensitivity to the mass ordering.

\section{Concluding remarks}
\label{concluding remarks}

This work presents a detailed analysis of mass ordering sensitivity using supernova neutrinos within the active-active and active-sterile frameworks, focusing in the context of the NO$\nu$A  experiment. We investigate the prospects of determining the neutrino mass ordering with the NO$\nu$A detector, considering the possibility that a supernova explosion might occur within the next five years. In such an event, NO$\nu$A could provide valuable insights into mass ordering sensitivity through supernova neutrino observations.  

This study also explores the impact of sterile neutrinos on mass ordering sensitivity. Specifically, if sterile neutrinos exist and were not present at the supernova core initially, but some active supernova neutrinos convert into sterile neutrinos before reaching the detector, we investigate how this transformation influences sensitivity. In the supernova neutrino simulation, for the first time, the neutral current (NC) channel is used to distinguish between active and sterile neutrinos. The difference in the total number of NC events directly explain the existence of sterile neutrinos.

The study also compares mass ordering sensitivity as a function of supernova distance for both active-active and active-sterile scenarios. The results show that the presence of sterile neutrinos enhances sensitivity for channels (iii) and (iv).  
Considering NC channel, although active-active scenario is blind on ordering conditions, there is a non-zero mass ordering sensitivity in active-sterile framework. When all events are combined, the NO$\nu$A experiment can achieve a 5$\sigma$ confidence level in distinguishing between normal and inverted mass orderings, assuming the supernova occurs at a distance of 5 kpc from Earth.
 The mass ordering coming from the NC channel is not that significant in the case of NO$\nu$A, but the efficient effect can be seen in future experiments with a large detector volume. 
The effect of systematic errors is also examined, demonstrating that the ``no-sys" condition yields the highest sensitivity values, while the inclusion of any systematic errors reduces sensitivity across all channels.  
Finally, the impact of energy smearing on sensitivity is explored. The results consistently show that non-zero energy resolution decreases sensitivity.

\section{Acknowledgments}
PP wants to acknowledge Prime Minister’s Research Fellows (PMRF) scheme
for financial support. RM would like to acknowledge University of Hyderabad IoE project grant no: RC1-20-012. We gratefully acknowledge the use of CMSD HPC facility of University of Hyderabad to carry out the computational works. 
We also acknowledge Dr. Samiran Roy for useful discussions.

\newpage 
\appendix
\renewcommand{\thetable}{A1}
 \section{Event rates for different channels}
\label{app-a}

\begin{table}[h!]
    \centering
    \begin{tabular}{||c||c||c||c||}
    \hline
    \hline
       Channel  &~ Framework: $3 \nu / (3+1) \nu$ ~ &  ~Ordering ~ & ~Event Number ~ \\
       \hline
       \hline
       Channel (i) (IBD)  &  $3 \nu$ & NO & 107404 \\
       \hline
          &  &  IO & 111099  \\
          \hline
          \hline
            &  $(3+1) \nu$  & NO & 106589 \\
            \hline 
              &    &  IO & 110255 \\
              \hline
       \hline
       Channel (ii) ($\bar{\nu}_e-^{12}\text{C})$  &  $3 \nu$  &  NO & 3456 \\
       \hline
          &  &  IO & 4007 \\
          \hline
          \hline
            &  $(3+1) \nu$  & NO & 3430 \\
            \hline 
              &    &  IO & 3976 \\
              \hline
       \hline
       Channel (iii) ($\nu_e-^{12}\text{C})$  &  $3 \nu$  &  NO & 2919 \\
       \hline
          &  &  IO & 2231 \\
          \hline
          \hline
            &  $(3+1) \nu$  & NO & 2890 \\
            \hline 
              &    &  IO &  2080 \\
              \hline
       \hline
       Channel (iv) ($\nu_e-e)$  &  $3 \nu$  &  NO & 310 \\
       \hline
          &  &  IO & 285 \\
          \hline
          \hline
            &  $(3+1) \nu$  & NO & 304\\
            \hline 
              &    &  IO & 219 \\
              \hline
              \hline
    Total NC   & $3\nu$  & NO  & 11845\\
    \hline
    &  &  IO & 11845\\
    \hline
    \hline
    & $(3+1)\nu$  &  NO  & 11475 \\
    \hline
    &  &  IO & 11100\\
    \hline
    \hline
    Combined events & $3 \nu$  &  NO & 125934 \\
    \hline
      &   &  IO &  129467 \\
      \hline
      \hline
        &  $(3+1) \nu$ & NO & 124688  \\
        \hline
          &  &  IO & 127630\\
          \hline
          \hline
    \end{tabular}
    \caption {
    Event numbers for different channels (Channel (i), Channel (ii), Channel (iii), Channel (iv)), NC Channel and all channels combined at a supernova distance of 1 kpc. NO (normal ordering) and IO (inverted ordering) represent the mass ordering, while $3 \nu$ [$(3+1) \nu$] represents the active-active [active-sterile] neutrino framework. We apply a minimum reconstructed energy threshold of 10 MeV in all our event rate computations.}
    \label{appen-1}
\end{table}


\bibliography{reference}

\end{document}